\documentclass[11pt]{article}
\usepackage[margin=1in]{geometry}

\title{A Hierarchy of Limitations in Machine Learning}
\author{Momin M. Malik\\
Berkman Klein Center for Internet \& Society at Harvard University\\
\href{mailto:momalik@cyber.harvard.edu}{momin\_malik@cyber.harvard.edu}}
\date{29 February 2020\footnote{Draft version 0.3.06. In submission. Please cite with link \url{https://arxiv.org/abs/2002.05193}.}}

\usepackage[round]{natbib}
\usepackage[dvipsnames]{xcolor}
\usepackage[colorlinks=true, linkcolor=Mahogany, urlcolor=magenta, citecolor=ForestGreen]{hyperref}

\defcitealias{student1931}{``Student,'' 1931}

\def\redact#1{#1}

\usepackage{url,lineno,microtype,subcaption}

\usepackage{amsmath,amsfonts,amssymb}

\usepackage{xfrac}
\usepackage{bbm}
\usepackage{bm}
\usepackage{upgreek}

\usepackage{tikz}
\usetikzlibrary{arrows,shapes}
\tikzstyle{varcircle}=[circle, inner sep=2pt, minimum size=25pt, draw=black]
\tikzstyle{textellipse}=[ellipse, minimum width=2cm, inner sep=0.25cm, draw=black, font=\fontsize{10}{10}\selectfont, node distance=1]
\tikzstyle{blank}=[rectangle, minimum size=1cm]

\usepackage{pgfplots}
\pgfdeclarelayer{background}
\pgfsetlayers{background,main}

\usepackage{forest}
\useforestlibrary{edges}

\usepackage{array}

\newcommand{\Err}{\mathrm{Err}}
\newcommand{\err}{\mathrm{err}}

\newcommand{\tr}{\mathop{\mathrm{tr}}}
\newcommand{\1}{\mathbf{1}}
\newcommand{\I}{\mathbf{I}}
\newcommand{\0}{\mathbf{0}}
\newcommand{\R}{\mathbb{R}}
\newcommand{\E}{\mathbb{E}}
\newcommand{\Var}{\mathop{\mathrm{Var}}}
\newcommand{\Cov}{\mathop{\mathrm{Cov}}}

\newcommand{\bH}{\mathbf{H}}

\newcommand{\X}{\mathbf{X}}

\newcommand{\x}{\mathbf{x}}
\newcommand{\y}{\mathbf{y}}

\newcommand{\cN}{\mathcal{N}}

\newcommand{\bbeta}{\bm{\beta}}
\newcommand{\bepsilon}{\bm{\varepsilon}}
\newcommand{\bmu}{\bm{\mu}}

\newcommand{\bSigma}{\bm{\Sigma}}

\begin{document}

\maketitle

\begin{abstract}
\noindent``All models are wrong, but some are useful,'' wrote George E. P. \citet{box1979}. Machine learning has focused on the \textit{usefulness} of probability models for prediction in social systems, but is only now coming to grips with the ways in which these models are \textit{wrong}---and the consequences of those shortcomings. This paper attempts a comprehensive, structured overview of the specific conceptual, procedural, and statistical limitations of models in machine learning when applied to society. Machine learning modelers themselves can use the described hierarchy to identify possible failure points and think through how to address them, and consumers of machine learning models can know what to question when confronted with the decision about if, where, and how to apply machine learning. The limitations go from commitments inherent in quantification itself, through to showing how unmodeled dependencies can lead to cross-validation being overly optimistic as a way of assessing model performance.
\end{abstract}

\section*{Introduction}

There is little argument about whether or not machine learning models are \textit{useful} for applying to social systems. But if we take seriously George Box's dictum, or indeed the even older one that ``the map is not the territory' \citep{korzybski1933}, then there has been comparatively less systematic attention paid within the field to how machine learning models are \textit{wrong} \citep{selbst2019} and seeing possible harms in that light. By ``wrong'' I do not mean in terms of making misclassifications, or even fitting over the `wrong' class of functions, but more fundamental mathematical/statistical assumptions, philosophical \citep[in the sense used by][]{abbott1988} commitments about how we represent the world, and sociological processes of how models interact with target phenomena. 

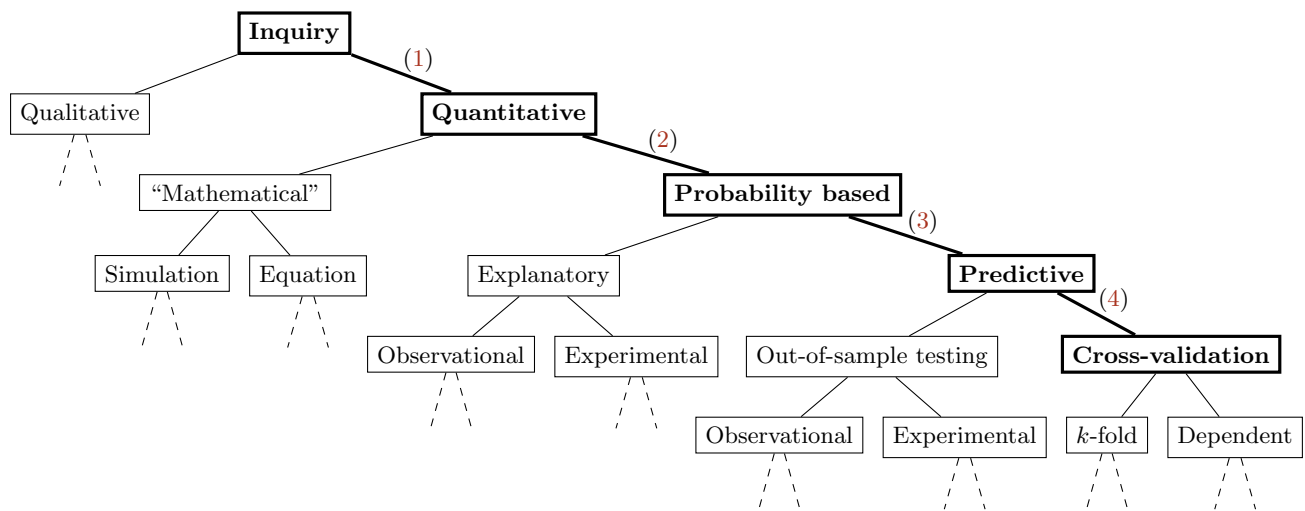
\begin{figure}[!ht]
\centering
\begin{forest}
for tree={draw, font=\footnotesize, minimum width=1em, minimum height=1em} 
[\textbf{Inquiry}, very thick, edge=very thick
	[Qualitative
		[, draw opacity=0, edge=dashed]
		[, draw opacity=0, edge=dashed]
	]
	[\textbf{Quantitative}, very thick, edge=very thick, edge label={node[midway,above right=-.33em,font=\footnotesize]{(\ref{sec:qual})}}
		[``Mathematical'' 
			[Simulation
				[, draw opacity=0, edge=dashed]
				[, draw opacity=0, edge=dashed]
			]
			[Equation
				[, draw opacity=0, edge=dashed]
				[, draw opacity=0, edge=dashed]
			]
		]
		[\textbf{Probability based}, very thick, edge=very thick, edge label={node[midway,above right=-.33em,font=\footnotesize]{(\ref{sec:model})}}
			[Explanatory 
				[Observational
					[, draw opacity=0, edge=dashed]
					[, draw opacity=0, edge=dashed]
				]
				[Experimental
					[, draw opacity=0, edge=dashed]
					[, draw opacity=0, edge=dashed]
				]
			]
			[\textbf{Predictive}, very thick, edge=very thick, edge label={node[midway,above right=-.33em,font=\footnotesize]{(\ref{sec:pred})}}
				[Out-of-sample testing
					[Observational
						[, draw opacity=0, edge=dashed]
						[, draw opacity=0, edge=dashed]
					]
					[Experimental
						[, draw opacity=0, edge=dashed]
						[, draw opacity=0, edge=dashed]
					]
				]
				[\textbf{Cross-validation}, very thick, edge=very thick, edge label={node[midway,above right=-.33em,font=\footnotesize]{(\ref{sec:CV})}}
					[$k$-fold
						[, draw opacity=0, edge=dashed]
						[, draw opacity=0, edge=dashed]
					]
					[Dependent
						[, draw opacity=0, edge=dashed]
						[, draw opacity=0, edge=dashed]
					]
				]
			]
		]
	]
]
\end{forest}
\caption{A diagram of the hierarchy explained in this paper, with each bold branch corresponding to a section.}\label{fig:tree}
\end{figure}

This paper takes a particular model of machine learning research or application: one that its creators and deployers think provides a reliable way of interacting with the social world (whether that is through understanding, or in making predictions) without any intent to cause harm \citep{mcquillan2018} and, in fact, a desire to not cause harm and instead improve the world,\footnote{I thank 
\redact{John Basl} for encouraging me to make clear that I consider both methodological and ethical limitations.} for example as most explicitly in the various ``\{Data [Science], Machine Learning, Artificial Intelligence\} for [Social] Good'' initiatives, and more widely in framings around ``fairness'' or ``ethics.'' I focus on the almost entirely statistical modern version of machine learning, rather than eclipsed older visions (see section \ref{sec:pred}). While many of the limitations I discuss apply to the use of machine learning in any domain, I focus on applications to the social world in order to explore the domain where limitations are strongest and stickiest. I consider limitations in machine learning such that, contrary to the expectations and intentions of creators and deployers, machine learning can fail to be reliable and/or can cause harm. I do so in a systematic review, structured along four decisions that are implicitly made when deciding to use machine learning:
\begin{enumerate}
    \item[\ref{sec:qual}.] To use quantitative analysis over qualitative analysis;
    \item[\ref{sec:model}.] To use probabilistic modeling over other mathematical modeling or simulation; 
    \item[\ref{sec:pred}.] To use predictive modeling over explanatory modeling;
    \item[\ref{sec:CV}.] To rely on cross-validation to evaluate of model performance.\footnote{Cross-validation is a method of model \textit{selection}, but also a method of model \textit{evaluation} \citep{hastie2009} by which to determine the validity of a machine learning claim. These two uses of cross-validation can have very different theoretical properties from each other \citep{wager2019}; I am concerned exclusively with the latter. Two other ways of evaluating model performance that I will consider are experiments and true out-of-sample testing, but cross-validation is the dominant method, at least in research papers.}
\end{enumerate}
This is conceptually illustrated in fig. (\ref{fig:tree}). None of these decisions, leading up to the choice of using machine learning and cross-validation, are inevitable; it is possible to \textit{not} use modeling, or machine learning, or any computer-based approach \citep{baumer2011}---and so these are the points at which we should consider consequences and weight against alternatives.

On the four decision points, I argue:
\begin{enumerate}
    \item[\ref{sec:qual}.] When choosing quantitative analysis, we sacrifice the ability to have narrative understandings of meaning-making, commit to working with proxies rather than the actual constructs of interest, and risk delegitimizing lived experience.
    \item[\ref{sec:model}.] When choosing probability-based modeling, we restrict our view to the world as consisting of fixed entities with properties, and prioritize the use of a central tendency over engaging with individuality. 
    \item[\ref{sec:pred}.] When prioritizing prediction, we make models that are not reliable for getting insight into the underling system, and are sensitive to changes and reactions.
    \item[\ref{sec:CV}.] When relying on cross validation, we risk being overly optimistic about a model's generalizability.
\end{enumerate}

The drawbacks of the four decision points form a hierarchy: the problems of prediction supersede all problems of cross-validation, the problems of the central tendency supersede all problems of prediction, and the problems of quantification supersede those of the central tendency. The hierarchy also be seen as a \textit{necessary chain of custody};\footnote{I thank \redact{Beau Sievers} for the term ``logical chain of custody.''} the results of cross-validation will only be as meaningful as the setup of a machine learning model, which will only be as meaningful as the way the world is put into observations and properties, which will only be as meaningful as way in which a phenomenon is quantified and measurement is done. Equivalently, limitations propagate through this hierarchy, with problems with quantification affecting everything downstream, and so on. 

I treat the levels in sequence, although interesting questions remain to ask relating these levels to each other, such as what limitations there may be in hermeneutic/interpretive (rather than modeling, especially based on central tendency amidst variability) approaches to data \cite[e.g.,][]{gibbs2013},\footnote{I thank \redact{Elena Fern\'{a}ndez Fern\'{a}ndez} for pointing me to this work.} or in dimension reduction techniques that see the need for qualitative interpretation of latent dimensions as a distinct strength \citep{desrosieres2012}.\footnote{I thank \redact{Nick Seaver} for pointing me to this work.}

\subsection*{Guide to reading this work}
The four sections of this work require progressively more technical background to follow, but they are also relatively self-contained. As a consequence, readers from different disciplinary backgrounds may refer only to specific sections covering topics of interest. 

Machine learning readers most interested in my technical argument of how to theorize the effect of dependencies on cross-validation should go to section (\ref{sec:CV}). Social scientists and policymakers considering whether and how to adopt machine learning (especially in place of social statistics) and the ethical implications of doing so should consider section (\ref{sec:pred}); a background in statistical modeling (notions of estimation, inference, model specification) are necessary for later parts of this section, which get into limitations of causal inference and experiments, but earlier parts are more historical, sociological, and philosophical. Section (\ref{sec:model}) identifies some frequently overlooked but fundamental assumptions of statistics and machine learning, and considers their consequences, which useful for all audiences. Section (\ref{sec:qual}) is a useful review for machine learning researchers about what qualitative research can do that machine learning cannot, and conversely, it can help qualitative researchers understand key gaps in machine learning and how qualitative methods can fill these.

\subsection*{Scope}
There are four important aspects of the scope. 

First, the limitations I identify are associated with using \textit{only} machine learning, which can therefore be seen as more a cultural problem among people who do machine learning than intrinsic limitations attributable to machine learning itself. But such machine learning-only approaches exist: for example, computer scientist Hanna \citet{wallach2018} talks candidly about her own process of realizing that ``machine learning is not a be-all and end-all solution.'' By using mixed methods, and including other approaches alongside machine learning for a given problem, many of these problems can be avoided.\footnote{I thank \redact{Berkman Klein Center fellows}, and particularly \redact{Apryl Williams} and \redact{Beau Sievers}, for encouraging me to make this point explicit.} That is: if machine learning is done with experimental validation via randomized control trials, then we will not suffer from over-optimism about performance. If we accompany predictive modeling with causal/statistical analysis, we can gain a better understanding of underlying processes, and better hope to make predictions that are robust to changes and reactions. And if we include qualitative analysis, we can understand constructs, and engage with individuality, narrative understandings, and lived experience. 

I do include several examples of mixed methods overcoming the limitations of machine learning alone \citep[e.g.,][respectively for validating lived experience and experimental validation]{patton2019a,cardoso2016}. But mixed methods research is not something that can done casually: it requires collaboration between researchers not only with different toolsets but potentially with vastly different understandings of the world, or else it requires individuals with multiple training in methods that individually can take a lifetime to master. So far, there is little work on on how to systematically do mixed methods machine learning research. Such work is needed to both precisely articulate what the difficulties are, and then understand how to overcome them. This is beyond the scope of this work, although in the conclusion I will return to this point as the most promising possibility for how to proceed.

Second, I do not mean to imply that alternative approaches, on their own, are superior to machine learning on its own \citep[][writes: ``Methods are like people: if you focus only on what they can't do, you will always be disappointed.'']{shapiro2014}. Each method has its own assumptions and resulting limitations, appropriate for different goals. The purpose of this paper is to highlight the specific limitations of \textit{machine learning}, and point out what are available methodological alternatives worth considering for any point at which machine learning may not be fit for purpose. 

So, if a researcher's goal is to understand meaning-making or identify underlying constructs, qualitative analysis is superior to machine learning, or indeed to any quantitative methodology. If the goal is to quantitively understand underlying data-generating processes \citep[especially in the midst of ``haphazard variability,'' and in a way that quantifies uncertainty;][]{cox1990}, statistics is superior to machine learning. If the goal is strong guarantees about generalizability, then experimental design or another validation methodology is superior to the use of cross-validation. But for pre-testing the design of a system for detection/classification by a given schema (especially if developed by or with relevant stakeholders), representing an arguably enormous space of use cases neglected prior to the development of machine learning \citep{breiman2001,mullainathan2017}, machine learning with cross-validation---for all its limitations---is very likely the appropriate approach. 

Third, for the most part I deliberately avoid discussing consideration around the \textit{data} that go into probabilistic models. The decision of what data to bring to a particular problem supersedes all other limitations and questions \citep{mallows1998}, but data has also been covered extensively, most recently around social media data in comprehensive review by \citet{olteanu2019}. The problems that can lead analysis astray include demographic sampling bias, limited data access, filtering decisions (especially selection on the dependent variable), and behavior driven by norms or constraints rather than ``naturalistic'' preferences. But there has been far less about properties of the probabilistic \textit{models} into which such data are fed, which this article seeks to address. Two aspects of data I do address, as they relate to what such models can do: what kinds of variability are captured in non-randomly sampled data, and the possibility of unobserved confounders. 

Last, this hierarchy is also limited to systematic inquiry. There are levels even higher up not included here, such as ways of knowing (which need not be systematic) and ways of being (which need not be structured around the pursuit of knowledge), as per Linda Tuhiwai \citet{smith2012}.\footnote{I thank Jonnie Penn for this point and reference.} That is, being a researcher or a developer is itself a choice that need not have been made; to borrow the language of Colin \citet{mallows1998}, that is the ``zeroth decision'' within this hierarchy. 

\subsection*{Purpose and contribution}
The paper draws together critiques from anthropology, sociology, science and technology studies, statistics, and machine learning itself, for the first time putting them into a comprehensive sequence. Then, the treatment of the effect of dependencies on biasing cross-validation using the idea of \textit{optimism} \citep{efron2004} is a simple but novel extension of this statistical theory that has the promise of being an elegant unifying framework for multiple disconnected efforts on how to think through properly structuring data splits. 

There are two intended audiences. First is machine learning modelers, to provide a comprehensive review of assumptions deeper than the mathematical ones made when doing modeling and intrinsic limitations induced by assumptions, as well as some examples of the assumptions breaking down to show concrete consequences. The second are audiences, consumers, and analysts of machine learning to better place the discipline among others, and to have a sense of contingencies they can consider when deciding if, where, and how to adopt machine learning. Critics in these audiences can hopefully use this to be \textit{specific} about their objections to machine learning, identifying the points at which they disagree with assumptions. 

Then, in addition to the purpose of serving the above two target audiences simultaneously (rather than make differentiated products), it is also to put social theory and statistical theory in conversation with each other. It agrees with Matteo \citet{pasquinelli2019}, who writes, ``the discussion about AI's limits may be inaccurate if technical limits are divorced from social limits, and vice versa,'' although as compared to that work (which addresses a non-expert audience and so spends time explaining core concepts and terms in machine learning), I take these for granted in addressing a machine learning audience and so can go far deeper into technical aspects. In methodological terms, this paper relates to ``\textit{in-situ} hybridization'' \citep{mackenzie2017}, treating different disciplinary forms of knowledge as coincident\footnote{I thank \redact{Nick Couldry} for reminding me to make explicit the link to social theory.} in order to relate the internal understandings of modeling to the larger social contexts of both the models and the understandings thereof. Unlike \textit{in-situ} hybridization, however, my approach is ultimately not an \textit{analytical} one, but an attempt to intervene as an actor \citep{collins2008}.\footnote{I thank \redact{Sheila Jasanoff} for pointing me to this work.} I advocate for specific interpretations and forgo interpretive flexibility, making this a work ultimately of practice and of a practitioner rather than of analysis. My work also relates to the call of Philip E. \citet{agre1997} for a \textit{critical technical practice} to add critical reflection to the development of technical knowledge, although this paper (in itself) lacks a praxis or demonstrated application and so ``critical technical theory'' might be more apt. Then, to preserve accessibility for different audiences, I do not pursue the same extent of co-incidence as Adrian \citet{mackenzie2017}, but divide the paper into modular sections over a \textit{hierarchy} of priorities. Mathematical portions are confined to the final section, and should be of particular interest to machine learning modelers to consider how to systematically think about cross-validation. 

The ideal effect of this paper paper would not simply be for those within machine learning to engage in methodological hedging---although being more humble about claim-making will be a positive outcome---or for those outside machine learning to retroactively justify an already-made decision to reject it. Even though mixed methods are not a part of this paper, by pointing out the limitations of machine learning as compared to other approaches, I hope to encourage cultural practices of putting machine learning in context with other possible approaches, and help develop systematic responses to limitations.\footnote{I thank \redact{Beau Sievers} for this framing.} This paper can be seen as a follow-up to \citet{selbst2019}: that work was about general limitations of mathematical abstractions, and this one is specifically about the structural and social limitations of the abstractions used in machine learning. I intend this to be a definitive work identifying and detailing these limitations that, going forward, can be the basis by which to plan out how to overcome these limitations and on which to structure efforts for doing so. 

\section{Quantitative versus qualitative analysis}\label{sec:qual}
Much has been written about how qualitative and quantitative methods are not mutually exclusive, and how they can complement and inform one another. This is certainly the case, although there are serious and deep philosophical conflicts \citep{bellotti2015,erikson2013}. But maintaining my focus on the branching paths that would lead to a machine learning-\textit{only} approach, here I focus on four failure points of a quantitative-\textit{only} approach. The first relates to the impossibility of quantifying meaning-making. The second is the difficulty of measurement in social science. The third relates to experience and personal knowledge. The fourth relates to how quantification can succeed by imposing its logic on the world, totally separate from any notion of correspondence or having empirical adequacy.\footnote{I am treating qualitative analysis here as something more systematic and specific than qualitative \textit{reasoning}; qualitative reasoning is universal, pervading every step of data analysis, such as picking which possible directions of data analysis to follow in what Andrew \citet{gelman2014} call the ``garden of forking paths.'' And, Donald \citet{campbell1975} points out that scientific results are uninterpretable without a ``narrative history'' portion that situates a work and the results. Peter \citet{spiegler2015} theorizes that mathematical modeling is a four-part process: of delimitation, denotation, solution, and interpretation, which involves two crosses across ``a significant linguistic divide'': delimitation goes from ordinary language to mathematical language, and interpretation moves back from mathematical language to ordinary language. Without these two moves, i.e. without both starting and ending in a realm where qualitative reasoning rules, modeling is meaningless. I thank \redact{Baobao Zhang} for this latter reference.}

\subsection{Meaning}
On the first point, qualitative researcher Michael Quinn \citet{patton2015}\footnote{I thank \redact{Maya Randolph} for pointing me to this work.} gives an argument for the ``nature, niche, value, and fruit of qualitative inquiry'':
\begin{quote}
    ``During the writing of this book, my first grandchild was born, and this book is dedicated to her. The hospital records document her weight, height, health, and \textit{Apgar} score -- activity (muscle tone), pulse, grimace (reflex response), appearance, and respiration. The mother's condition, length of labor, time of birth, and hospital stay are all documented... But nowhere in the hospital records will you find anything about what the birth of Calla Quinn \textit{means}. Her name is recorded but not why it was chosen by her parents and what it means to them. Her existence is documented but not what she means to our family, what decision-making process led up to her birth, the experience and meaning of the pregnancy, the family experience of the birth process, and the familial, social, cultural, political, and economic context that is essential to understanding what her birth means to family and friends in this time and place. A qualitative case study of Calla's birth would capture and interpret the story and meaning of her entry into the world from the perspectives of those involved in and touched by her coming into our lives.''
\end{quote}

What Patton describes here is ``thick description'': including the specificity and larger context, and getting at \textit{meanings} behind behavior and expressions. Theodore \citet{porter2012} links the converse, ``thin description,'' to the kind of descriptions provided by quantification and modeling: ``In a thin world, surfaces should be valid and deeper meanings superfluous.''\footnote{Here, in modeling terms, I would not take ``surface'' to be synonymous with \textit{observables}: while latent variables can be a way to express deeper meanings than what is directly observed, they might also only go only one layer deeper than the surface and so not be much less superficial.} He sees anthropologist Clifford Geertz's classic championing of ``thick description'' as ``a battle cry against an idea of social science harmonized, if not unified, by a shared commitment to the external and observable.''\footnote{I thank \redact{Rodrigo Ochigame} for pointing me to this piece.} Quantitative inquiry requires standardization and generalization; like any abstraction, this can succeed at its aims \citep{stinchcombe2001} and be seductive in its power \citep{selbst2019}, in what \citet{porter2012} calls the ``siren song of thinness,'' where ``the sacrifice of human meaning seem[s]... a price worth paying for solid results.'' 

In referencing what his granddaughter's birth \textit{means}, Patton's quote also connects with the concept of \textit{meaning-making}. This concept, originating from psychiatrist and holocaust survivor Viktor Frankl and also connecting to the sociological tradition of symbolic interactionism, is about the linguistic categories through which people make up their reality and define, justify, and interpret their actions. These are theorized as the most fundamental aspect of human social settings \citep{krauss2005}. Choosing quantitative analysis gives solidity but precludes the study of meaning-making---even as quantification relies on existing meanings. 

This is one standard by which qualitative inquiry is superior to quantitative: quantitative inquiry is \textit{unable to account for itself}. Quantitative analysis is also at the mercy of the quantification process; all exploration or discovery is strictly limited to variables that have been measured, whereas qualitative analysis has no such constraints \citep{campbell1975}. 

While, again, I am not going into the drawbacks of alternatives to the path that leads to machine learning, here it is worth noting that the quote above points to some limitations of qualitative analysis: as \citet{patton2015} writes, the ``physiological and institutional metrics'' around birthing ``provide trend data about the beginning of life... when aggregated across many babies and mothers.'' If our goal is such trend data, then measurements such as the \textit{Agpar} score are appropriate, while the meaning of Patton's granddaughter Calla Quinn's birth is not. In terms of drawbacks, while the ``subjective'' and contingent nature of qualitative inquiry should not be seen as a drawback since the studied phenomena are themselves contingent, the usual objections of a lack of scalability and generalizability apply. 

Furthermore, qualitative inquiry is not necessarily more ethical. In practice, qualitative inquiry has developed strong ethical frameworks \citep{morethancode2018}, but this has been in response to both historic and contemporary failures to commit to principles of reciprocity and sharing knowledge that Linda Tuhiwai \citet{smith2012} argues are necessary for non-exploitative research. After all, exposing the inner meanings that people hold can be more invasive than any surveillance, and indeed, ethnography and ``thick'' descriptions have both historically played and continue today to play a role in furthering oppression and domination \citep{smith2012,price2011}.\footnote{Again, I thank \redact{Rodrigo Ochigame} for pushing me to this point.} In another critique, medical anthropologists \citet{kleinman1995} write how ethnographers frequently treat human suffering with detached analysis which is ``every bit as dehumanizing as that of [medical] colleagues who unreflectively draw upon the tropes of tropical medicine or behaviorism to create their subject matter.''

Similarly, the principle of reflexivity---self-awareness by researchers about both their effect on the research, and the effect of the research on them \citep{attia2017}, which is not just a subjective process but an \textit{intersubjective} one that considers relationships and interconnectedness between different people's subjectivities \citep{cunliffe2016}\footnote{I thank Maya Randolph for this reference.}---does not guarantee responsible research. One can be perfectly aware of being an exploiter. Nor is it impossible to be reflexive in quantitative research. However, it is considerably harder when working with layers of mathematical forms that require years of specialized training to understand and internalize as natural and intuitive, abstractions not shared by the potential subjects of study. Perhaps this distance makes the need for reciprocity even more pressing: Dan \citet{mcquillan2018} suggests that, because ``Machine learning represents one of the highest historical forms of the abstraction of social relationships,'' it ``needs to be counterbalanced by the unmediated relationships of popular assemblies.''

\subsection{Measurement}
Second is the question of measurement. Qualitative research is able to get more directly at processes of interest, engaging directly with how they are multifaceted, variable, and context-dependent. But when moving to quantification, what is measurable is almost never what is of interest in social systems \citep{devellis2017}; in 1939, John Maynard \citet{keynes1939} expressed skepticism towards mathematical modeling in early econometrics partly based on the difficulty of measuring ``political, social and psychological factors, including such things as government policy, the progress of invention and the state of expectation.'' John Stuart Mill had similar concerns even earlier, in 1871, about mathematics in economics, which Peter \citet{spiegler2015} summarizes as the worry that ``mathematical language might generate a purely quantitative conceptual map of the subject matter it purported to outline, with no way of telling whether the outlines on the map corresponded to the subject's own contours.''

No quantification captures everything; like modeling more generally, a measurement is a reduction that solidifies some meanings while excluding others. Yarden \citet{katz2017} explains how machine learning (and the ``connectionist'' tradition in artificial intelligence) echoes the old behavioralist paradigm in a focus ``on an input-output relationship that's learned by tuning the model using error signals in data, without worrying about any internal states that govern that relationship.'' This prioritization is a bargain that machine learning makes; it has achieved successes far beyond those of previous behavioralist research, but this section is about where, how, and for what goals this bargain can fail. 

Psychometrics, in particular, has developed extensive theory around the idea of an underlying \textit{construct}, which is the entity of actual interest, versus the proxies used to measure it \citep[although these efforts are not without serious problems, e.g.,][]{michell1999}.  ``Constructs'' are things like well-being \citep{alexandrova2005}, user engagement, creditworthiness, or `intelligence,' which we attempt to measure, respectively, with proxies like self-report, clickthrough rates, correlates of loan repayments, and standardized test responses.\footnote{This is not to say that we should treat every abstract social idea as a construct; we could distinguish constructs from, say, values. For example, for the World Press Freedom Index, we could say that ``press freedom'' is a construct which we can measure by proxy. But it would be more precise to identify the construct as a normative standard of ``how professional journalists should act and exist in society,'' and the goal to be to measure the obstacles to them meeting this standard. I thank \redact{Joshua Kroll} for a critique of treating values as constructs, from where I got this point.} Even ``crime'' is a construct, insofar as (even given a particular legal code \textit{defining} crime) there are no measures of crime: there are only data of arrest records, or of crime reports, or of incarceration, none of which is one-to-one with actual crime \citep{ochigame2018,elliott1995}. \citet{pierce2008} analogizes proxies to asking, ``Can we really weigh an iceberg by measuring its tip?'' If quantitative research does not deal with the problems of proxies, it is akin to thinking that the tip of the iceberg is its entirety. And, following the analogy, predicting to what is available is akin to predicting to the tip of the iceberg. Sometimes this will be sufficient, but many times, it will not be. 

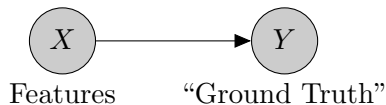
\begin{figure}[!ht]
\centering
\begin{tikzpicture}[>=latex,text height=1.5ex,text depth=0.25ex]
	\matrix[row sep=1.0cm, column sep=.6cm] { 
	\node (x) [varcircle,
			fill=black!20,
			label=270:Features] {$X$}; &
	\node (y) [varcircle, 
			fill=black!20,
			label=270:``Ground Truth''] {$Y$} ; \\
	};

	\path[->, >={triangle 45}]
	(x) edge (y)
	; 
\end{tikzpicture}
\caption{A graphical model \citep{pearl2009} that gives the conceptual setup of a typical supervised learning task. Nodes in gray are observed, and nodes in white (not present here) are unobserved, and arrows show causal directionality. Generally, we imagine that the features, $X$, ``produce'' or cause the ``ground truth'' label, $Y$, but the causal direction is not important for a machine learning application; for prediction (see section \ref{sec:pred}), the only relevant factor is that the two are correlated. This can happen identically whether it is $X$ that causes $Y$, $Y$ that causes $X$ (as in a detection task), or an unobserved common cause that causes both $X$ and $Y$.}\label{fig:xy}
\end{figure}

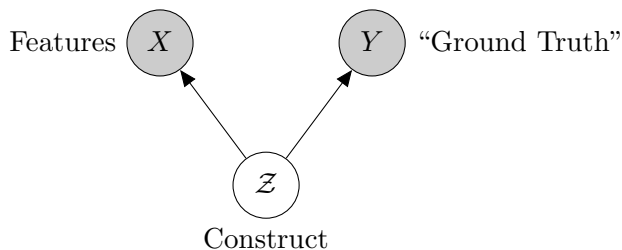
\begin{figure}[!ht]
\centering
\begin{tikzpicture}[>=latex,text height=1.5ex,text depth=0.25ex]
	\matrix[row sep=1.0cm, column sep=0cm] { 
	\node (x) [varcircle,
			fill=black!20,
			label=180:Features] {$X$}; & &
	\node (y) [varcircle, 
			fill=black!20,
			label=0:``Ground Truth''] {$Y$} ; \\
	& \node (z) [varcircle, 
			label=270:Construct] {$\mathcal{Z}$}; \\
	};

	\path[->, >={triangle 45}]
	(z) edge (x)
	(z) edge (y)
	; 
\end{tikzpicture}
\caption{A graphical model showing the place of a construct in the supervised learning task illustrated in fig. (\ref{fig:xy}). There is an unobserved, potentially unmeasurable, underlying construct $\mathcal{Z}$ that produces both the labels $Y$ and the associated features $X$, such that $X$ and $Y$ are correlated. }\label{fig:xyz}
\end{figure}

Less metaphorically, constructs (the ``inner states'') can be formulated in a latent variable model \citep{devellis2017}; that is, not every latent variable represents a construct, but constructs can be expressed as latent variables. If, conceptually, the typical supervised learning task looks like fig. (\ref{fig:xy}), then the place of underlying construct can be illustrated as in fig. (\ref{fig:xyz}). Unlike ``ground truth,'' which is something measurable, a construct (which is closer to an actual underlying ``truth'') is something latent and potentially unmeasurable but that produces observable measures (with some noise, and/or with other complicating inputs). We can potentially use this to estimate latent values. 

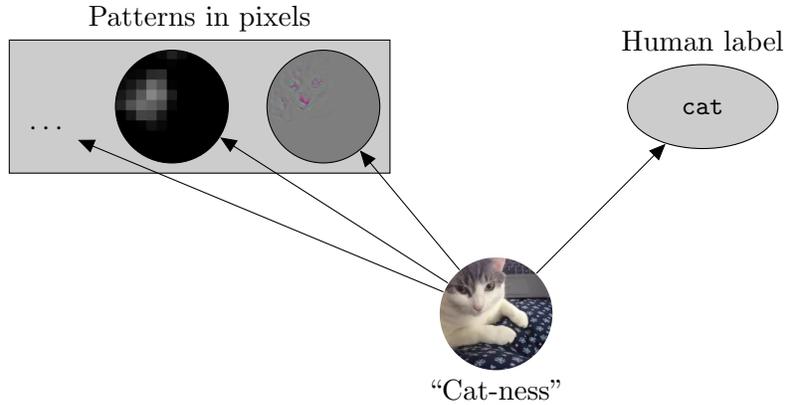
\begin{figure}[!ht]
\centering
\begin{tikzpicture}[>=latex,text height=1.5ex,text depth=0.25ex]
	\matrix[row sep=1.25cm, column sep=.5cm] {
	\node (blank) [anchor=north] {$\cdots$}; &
	\node (pixel1) [circle, anchor=center, minimum size=1.5cm, draw,
		path picture={\node at (path picture bounding box.south){
	           	\includegraphics[width=1.75cm]{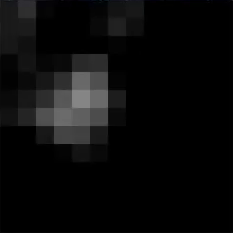}
		};
		}]{}; &
	\node (pixel2) [circle, anchor=center, minimum size=1.5cm, draw,
		path picture={\node at (path picture bounding box.south){
	           	\includegraphics[width=1.75cm]{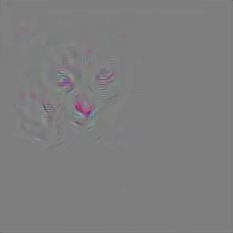}
		};
		}]{}; & &
	\node (label) [textellipse, align=center, fill=black!20,
				label=90:Human label] {\texttt{cat}} ; \\
	& & & 	\node (cat) [circle, anchor=center, minimum size=1.5cm, label=270:``Cat-ness'', 
		path picture={\node at (path picture bounding box.south){
	           	\includegraphics[width=1.75cm]{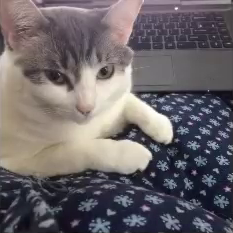}
		};
		}]{}; & \\
	};

	\path[->, >={triangle 45}]
	(cat) edge (label)
	(cat) edge (blank)
	(cat) edge (pixel1)
	(cat) edge (pixel2)
	;
	\begin{pgfonlayer}{background}
		\node [rectangle, draw=black, fill=black!20, fit=(blank) (pixel1) (pixel2), label=90:Patterns in pixels] {};
     \end{pgfonlayer}
	
\end{tikzpicture}
\caption{The way in which a construct underlies the specific task of image recognition. Images from \citet{yosinski2015}, used under a CC BY-NC-SA 3.0 license.}\label{fig:construct}
\end{figure}

To take a more concrete example, imagine a supervised learning task to recognize an image as being of a cat or not being of a cat. Here, the underlying construct is something we might call ``cat-ness'' (fig. \ref{fig:construct}), which is not directly measured; what is available is the \textit{human labeling of images} as being of cats or not of cats. 

\begin{figure}[!ht]
\centering
\includegraphics[width=2in]{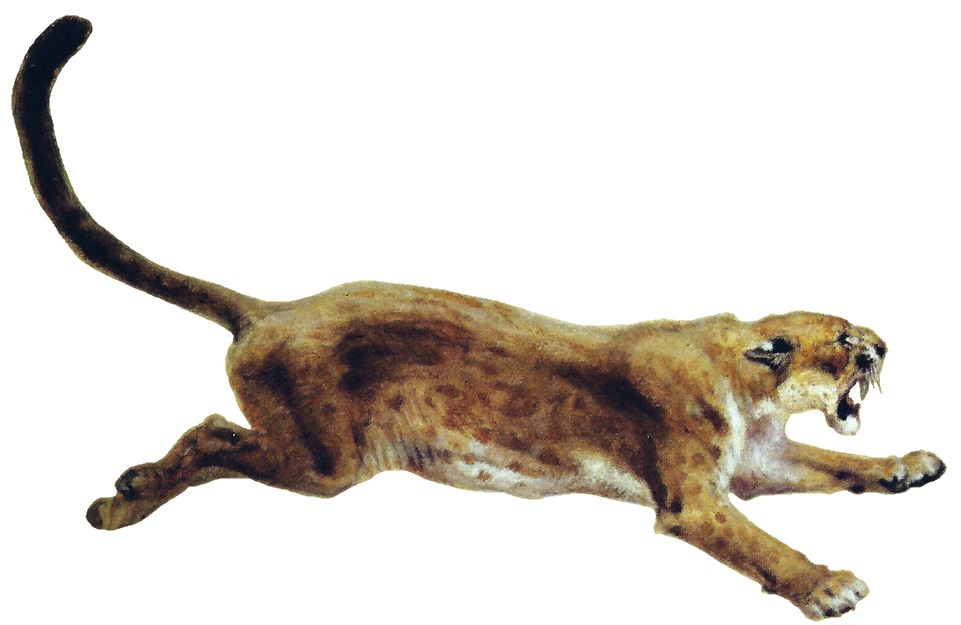}\hspace{.5in}\includegraphics[width=2in]{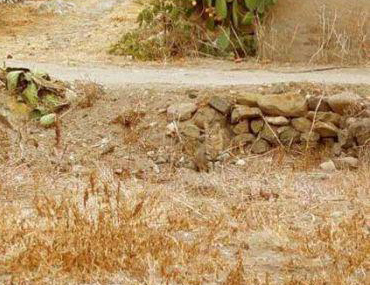}
\caption{Left: not an image of a cat, but potentially possessing cat-ness. Right: an image of a cat, but potentially lacking cat-ness.}\label{fig:cat}
\end{figure}

That is, what is used as ``ground truth'' is not cat-ness itself; rather, cat-ness both produces both the human labels, and the groupings of pixels (which is why the two are correlated). As far as we currently know, the specific ``patterns of pixels'' identified by neural nets are quite far from how human perception works, even if there are overlaps \citep{zhou2019}. But also, the construct cat-ness is not some Platonic form; we have to decide what we want a notion of underlying cat-ness to accomplish. Depending on our goal, even human perception is not a foolproof proxy. Perhaps we might want to identify cat-ness with phylogeny, of being of the family Felidae; in such a case, the left image in fig. (\ref{fig:cat}) lacks cat-ness, since it is Charles R. Knight's 1904 illustration of Dinictis (a genus of the extinct family Nimravidae, known as ``false saber-tooth cats'') and not a cat. Alternatively, we might want to define the cat-ness of an image as whether or not people perceive a cat; in that case, the image on the left of fig. (\ref{fig:cat}) would possess cat-ness, whereas the image of the right of fig. (\ref{fig:cat}) would lack it.\footnote{We should, however, be careful to specify who we mean by ``people,'' as people who are primed with the information that there is a camouflaged cat in the picture \citep{rosenfeld2018} and who are determined enough will find the cat.}

As we can imagine from this example, it is an extremely difficult undertaking to decide how to set up the hypothesized latent structure in order to estimate or even quantify latent values, decide what observable measures to use for which hypothesized latent entities, and decide how to determine success of inference to something unobservable. 

My focus here is on limitations of not \textit{accounting} for constructs; for a comprehensive overview of how machine learning can use ideas of constructs and validation, I point readers to \citet{jacobs2019}. But briefly, psychometrics validates proposed models relating measurements to constructs multiple ways. For example:
\begin{itemize}
\item External validity is whether associations seen in the study generalize beyond it; it is synonymous with generalizability; I discuss it further in section (\ref{sec:pred}) as the sole focus of machine learning. 
\item Criterion-related validity\footnote{This is also called ``predictive'' validity, but I uses criterion-related to distinguish it from ``prediction'' in the sense of machine learning, which is about external validity.} is when a measure should predict some auxiliary outcome that the construct is hypothesized to cause (e.g., a measure of well-being should, say, correlate with health indicators, preferably before those indicators are manifested if there is a theoretical model that well-being causes health). 
\item Internal validity is whether conclusions about causal relationships within the study generalize. 
\item Content validity relates to whether the measurement can cover the total possible range of the construct (or effectively sample within it). 
\item Face validity is if something makes sense based on known relationships. So, for example, a quantitative model that can show that patterns in tea leaves in a cup of hot water predict real-world events would lack face validity, even if it succeeds empirically. 
\end{itemize}

As uncertain as this process, is, \textit{any quantitative research that ignores it can only make pronouncements about what is directly measurable}. Or, when the goal is not study but simply to construct systems, ``concept drift'' \citep{schlimmer1986} may be understood as a consequence of using what is measurable rather than identifying the underlying construct. The ``hidden context'' is the actual target, rather than a sort of complicating factor to be managed \citep{widmer1996} using measurables. 

For example, if not considering issues of measurement, studies of online social networks are only able to say something about ties on those networks and not of the construct of `friendship' that those ties supposedly capture. Or, if there are untrained annotators applying their own heuristic understandings to produce labels, the results may be meaningless and useless \citep{patton2019a} regardless of how how high the ``accuracy'' is of a model to recover those annotations. 

This also relates to what David Hackett \citet{fischer1970} names as ``quantitative fallacy'': relying on what is measurable to the exclusion of all else. This was also later named the ``McNamara fallacy'' after the disastrous reliance of the US Secretary of Defense during the Vietnam War, Robert McNamara, solely on reported measures of enemy body counts as his measure of success \citep{omahony2017} without even any validation. This gave incentives to those reporting to him to simply lie about the numbers (see also Sec. \ref{sec:gaming}). A recent result by \citet{obermeyer2019} can be interpreted in terms of constructs: they detail a machine learning model used by a major healthcare company that de-prioritized care for black patients versus equivalently sick non-black patients. This was because model optimized to healthcare \textit{costs}, which are easy to measure, but the actual target of interest is the hard to define and measure construct of \textit{illness}. Costs are not a solely a proxy solely of illness; they also capture structural racism, as less money is spent caring for Black patients. Consequently, the use of what is available and measurable led to a system that formalizes discrimination against Black patients. 

Arguably, machine learning can defer responsibility for any problems with quantification to whatever is given to it as the ``ground truth.'' That is often fair as a division of labor. But many times in practice, when machine learning makes claims, it forgets to recognize and communicate that the measurements serving as ``ground truth'' are a black box that hide problems of measurement, and that the ground truth is not construct itself. This is especially true when, as documented by \citet{geiger2020} for social computing, machine learning researchers frequently report doing the labeling themselves, or else do not even report where labels come from. At other times (e.g., when a machine learning researcher is working with haphazardly collected or legacy data, or when the data collectors are themselves untrained in and/or irresponsible about measurement, constructs, and validation), there may be nobody to defer the responsibility to; in such cases, if a machine learning researcher wants their research to be meaningful and effective, they must themselves take up responsibility of the quantification process. Or, at the very least, they should maintain some awareness of the contingencies and uncertainties of quantification and measurement, and communicate appropriate caveats. 

Measurement can also be mutli-layered, with high-level problems compounding on underlying ones. Consider Google Ngrams.\footnote{I thank J\"{u}rgen Pfeffer for this example.} First, those $n$-grams are built from a measurement technique: scans run through optical character recognition (OCR). For over a hundred years' worth of data in the corpus, lowercase `s' was typeset in a way that made OCR classify it as `f' \citep{wired2015}. Second, $n$-grams are a frequency measurement, but as \citet{pechenick2015} argue, these frequencies do not work as a measure of popularity: the Google Books corpus contains one copy of each book, which equally weights blockbuster hits along with books that never sold a single copy. One of the original applications of n-grams was as a measure of censorship in Nazi Germany; but \citet{koplenig2015} documents how the lack of metadata make it questionable whether the results reflect censorship or are even linked to World War II. Such measurement problems are over and above problems coming from an unclear sampling frame \citep{pechenick2015}. Consequently, \citet{koplenig2015} recommends we ``explicitly restrict the results to linguistic or cultural change `as it is represented in the Google Ngram data';'' but such a total lack of generalizability was hardly the intention of the corpus. 

\subsection{Experience and legitimacy}
One last consequence of quantitative analysis is that it centralizes power in the analyst. This is not intrinsically bad, but can enable oppressive outcomes. And, this can happen in qualitative analysis as well, but for those methodologies there exist methods of co-creation, and specifically, participatory action research \citep{morethancode2018} that seek to avoid this. 

With quantification locking in one possible set of meanings, a consequence is the risk of delegitimizing the experiences of individuals. For example, why should quantitative evidence \cite[e.g.,][]{gelman2007}, rather than people's lived experience, be the arbiter of whether or not policing is biased? In a classic example of Simpson's paradox---graduate admissions at UC Berkeley---it was determined that, after controlling for departments, there was no evidence of bias towards men \citep{bickel1975}. But the authors of this analysis recognized that this required a specific assumption, that men and women had equal academic qualifications, and that the real process of interest, discrimination, is the \textit{unobservable} ``exercise of decision influenced by the sex of the applicant when that is immaterial to the qualifications for entry.'' Even if internal decision-making of admissions officers could be observed, as Deborah \citet{hellman2008} also points out, the criteria for defining ``merit'' are themselves subjective and endlessly contingent (how would we compare one person to another who worked twice as hard to achieve, say, just less than half as much?) and could be discriminatory. Contrast this to a possible qualitative approach, not considered by the authors, which would be to examine the \textit{experience} of women in the graduate programs and their differential treatment, such as in a study from shortly after by \citet{sandler1986}. 

Structures of power may commit to using quantitative evidence as a neutral ground. Then, just as Gayatri \citet{spivak1985} theorizes about ``strategic essentialism''---the adoption of essentializing identities by marginalized peoples as a tactic for gaining recognition---we can discuss ``strategic quantification,'' or ``strategic modeling.''\footnote{I thank \redact{Os Keyes} for making this connection between strategic essentialism and data/modeling.} 

Quantitative evidence may also be a tactic for majority groups to combat their own prejudice \citep{wise2019}. And \citet{abebe2020} theorize that computing (i.e., modeling and quantitative approaches) can be socially useful for tasks like diagnosis/measurement, formalization, or rebuttal. As an example of ``rebuttal,'' the Clark Doll Experiments of the 1940s were cited as evidence about the psychological harms of segregation in the Brown v. Board of Education case (rebutting claims of segregation being benign or beneficial), in a famous example of quantitative analysis successfully contributing to real change \citep{henry2004}. However, this is perhaps an exception. While we can easily find hundreds, perhaps even thousands, of works quantitatively ``proving'' some instance of inequality, how many have led to actual change? Ruha \citet{benjamin2019} notes that there is a long history \citep{wells1895,dubois1899,dubois1900,battle2018} of making suffering and injustice quantitatively visible, but such efforts have often proved ineffective. She cites experimental evidence \citep{hetey2018} suggesting efforts can even be counterproductive---that exposure to to quantitative evidence of extreme disparities can create \textit{more} support for the policies that create those disparities among those who do not experience them---and herself advocates \textit{narrative tools} as the more effective strategy for change. That is, perhaps narrative tools are both necessary and sufficient, whereas quantitative approaches are neither necessary nor sufficient. 

When not chosen strategically but instead imposed, insisting that people from marginalized groups quantitatively prove their experiences---especially when this demand is made only to members of those groups---is unjust \citep{lanius2015}. And effort spent towards doing this also falls under a critique by Toni \citet{morrison1975} pointing out the perversity and fecundity of quantitative (and other) standards:\footnote{\redact{Maya Randolph} connected this statement to efforts quantitatively, statistically ``prove'' racism.}
\begin{quote}
``The function, the very serious function of racism, is distraction. It keeps you from doing your work. It keeps you explaining over and over again, your reason for being. Somebody says you have no language and so you spend 20 years proving that you do. Somebody says your head isn't shaped properly so you have scientists working on the fact that it is. Somebody says that you have no art so you dredge that up. Somebody says that you have no kingdoms and so you dredge that up. None of that is necessary. There will always be one more thing.''
\end{quote}

Indeed, when model outcomes do not benefit structures of power, those structures can, and do, called the modeling into question. One example is a reply from advocates of the auto lending industry to a report from the Consumer Financial Protection Bureau (CFPB) statistically arguing for existence of discrimination by indirect auto lenders, and citing those lenders for fines \citep{americanbanker2015}. The reply calls the CFPB methodology (specifically, inferring unobserved traits using proxies) ``essentially sophisticated guesses'': but so much modeling, including the modeling used by the cited industry to benefit itself, could be called the same. 

Harry \citet{collins1981} introduced the idea of ``experimenter's regress'' to discuss how the outcome of experiments depend on proper apparatus and experimenter competence, such that it is always possible to challenge experimental results \citep{mackenzie1989}. It is similar to the earlier idea of \textit{confirmation holism}, where all claims rely on supporting claims that can also be challenged (and so multiple claims have to be accepted simultaneously even if they are logically sequential), and has also been extended to the idea of theoretician's regress \citep{kennefick2000}.\footnote{I thank \redact{Nick Seaver} for this connection and the extension.} We can similarly extend this to choices around probabilistic modeling and propose ``modeler's regress''---which is how the limitations and uncertainties of modeling can always be used to strategically dispute the legitimacy of a particular model when its conclusions are unfavorable. R. A. Fisher himself, an avid smoker and paid consultant for the tobacco industry, ended his life trading in his authority to vigorously argue against the emerging consensus about the causal link between smoking and lung cancer \citep{stolley1991}. The essential variability of the modeling process itself is on display in the amazing study of \citet{silberzahn2018}, where twenty-nine teams of statistical analysts given the same data set and the same target quantity to estimate, and not a single team used the same statistical model! Reassuringly, the confidence intervals of most of the studies overlapped, giving an overall robust result, but this shows how enormously variable the modeling process is and how much room this suggests there is for making challenges. 

What are ways to mitigate the limitations of quantitative-only research? One is to center narrative, rather than quantitative tools, when trying to carry out advocacy. Another is that, for quantitative-only research or quantitative components, any claims about unmeasurable constructs based only on analysis of measurable proxies should recognize the uncertainty (whether quantitatively, such as in terms of measurement error, or otherwise) that comes from this distance. Most ambitious and promising, though, is to make qualitative inquiry an integral part of any rigorous research project. For example, developing labeling schemes is an area where qualitative researchers have extensive training and experience and which can entirely change the results of a machine learning classification task. \citet{frey2018} and \citet{patton2019a,patton2019b} argue for this; but these works also embody something even more important, which is for qualitative research can frame overall projects. That way, qualitative research can also determine the meaningfulness of final quantitative results to the actual systems; indeed, \citet{wagstaff2012} argues that even for a machine learning model to succeed on accuracy, or precision, or any such abstract metric ``tells us \textit{nothing at all useful} about generalization or impact.'' A qualitative analysis of a quantitative system, if it can be properly integrated into an overall project, may tell us much more.

\subsection{Performativity}
A pragmatic objection is to say: so long as we can reliably anticipate system outputs, do the limitations of the sorts outlined above really matter? The response is to emphasize again that it may be perfectly possible to build a model that anticipates with high accuracy, precision, and recall how an ignorant annotator would label a given instance, but such modeling will not provide meaningful statement about the system or underlying constructs.

But further, what if we use a model outputs to successfully take action? Would this not show that the limitations do not seriously affect modeling? In response, we must consider the \textit{performativity} critique: that models do not work because they are ``true,'' but because they (or rather, their creators and the loci of power that deploy modeling) \textit{impose the logic of the models on the world}.\footnote{More precisely, Kieran \citet{healy2015} summarizes: ``Originally articulated by \citet{callon1998} and refined by \citet{mackenzie2003}, and \citet{mackenzie2006}, the \textit{performativity thesis} is that economics produces a body of formal models and transportable techniques that, when carried out into the world by its professionals and popularizers, reformats and reorganizes the phenomena the models purport to describe... The success of economics is not just a matter of a particular conception of rationality serving as a ceremonial gloss on social action; nor is it a simple instance of ideological indoctrination... Rather, tools implementing formal models of action---`calculative devices'---are put in the hands of social agents by the model-builders or their representatives. These devices act as `cognitive prostheses' that enable actors to accomplish calculative tasks previously beyond their reach, but which are required by the theoretical models. When incorporated into the everyday work of market agents, these devices allow real settings to better approximate the original models, and their assumptions.'' I thank \redact{Abby Jacobs} for pointing me to this work.} This critique was originally developed around economic models, where it is easy to see how, once people begin optimizing to something like the Black-Scholes model, it becomes true regardless of whether or not it successfully described the system beforehand. But it applies much more widely \citep{healy2015}, including open opportunities to apply to machine learning---even if, unlike the strong claims of economics \citep{syll2018}, machine learning is generally agnostic about mechanisms. 

While there are clear-cut empirical cases of a model working by transforming the world, rather than necessarily by reflecting it \citep{malik2016b,malik2018}, a strong form of the performativity critique is to say that models \textit{only} work by transforming the world. An intermediate form of performativity is that people react to being modeled, creating nonstationary environments and leading to concept drift that is potentially what makes the model work.\footnote{Thanks to \redact{Joshua Simons} for suggesting the connection of concept drift to performativity. Connecting to the earlier mention of concept drift around constructs, here we would say that there is some underling, stationary but probably unmeasurable, construct consisting of how people react to certain incentives.} As Nicholas \citet{rescher1998} notes that in cases of feedback loops, it is not actually the model itself, but the \textit{reaction} to it that can become causal and makes predictions self-fulfilling. And a light form of performativity is to note that self-driving cars have not exactly been successful in navigating an open-ended social and natural environment: rather, they require both the social and natural aspects of the environment to be manipulated to succeed. The ultimate success of integrating fully autonomous cars alongside human drivers will likely require training human drivers as well. Thus, even if not committing to the strongest form, performativity can be a powerful analytic frame for generating hypotheses about sources of modeling success that can be empirically investigated (including through modeling itself). 

\section{Probability-based modeling}\label{sec:model}
When it comes to social systems, the probability-based modeling of statistics and machine learning is far more popular that other approaches. It has two attractive features---namely, being able to account for variability and being able to make use of data, properties that competitors do not have. However, it requires making commitments in terms how it structures the world, and how it reduces data to usable information.

\subsection{The data matrix}
In a critique of regression modeling in social science, Andrew \citet{abbott1988} identified how such modeling assumes the world can be divided into entities (the observations, or instances), and properties of those entities (the variables, or features). He continues,
\begin{quote}
``...it is striking how absolutely these assumptions contradict those of the major theoretical traditions of sociology. Symbolic interactionism rejects the assumption of fixed entities and makes the meaning of a given occurrence depend on its location -- within an interaction, within an actor's biography, within a sequence of events. Both the Marxian and Weberian traditions deny explicitly that a given property of a social actor has one and only one set of causal implications. Marx's dialectical causality makes events produce an opposite as well as a direct outcome, while Weber and the various hermeneutic schools treat attributes as infinitely nuanced and ambiguous. Marx, Weber, and work deriving from them in historical sociology all approach social causality in terms of stories, rather than in terms of variable attributes.''
\end{quote}
That is, the very formulation of the world into a data matrix is neither inevitable nor natural. In philosophical terms, the data matrix either makes the \textit{ontological} assumption that the world is indeed structured as entities and properties of those entities, or it makes the \textit{epistemic} assumption that the world is meaningfully knowable/approximated in such terms.\footnote{I thank \redact{Sabelo Mhlambi} for noting the need to make the philosophical angle explicit.}

Especially if we internalize these assumptions and forget that they are, indeed, assumptions, they limit what we can imagine and describe. In terms of power and control, Mark \citet{poster1995} wrote about how databases (whose tables are equivalent to data matrices) fragment people by ``grids of specification,'' both multiplying and decentering them, which allows them to be recalled and reassembled for the purposes of the database owners and users.\footnote{This reference is from \citet{raval2019}.} In inheriting this formulation of the world, machine learning creates the same sorts of reconfigurations of subjects \citep{mackenzie2017}. 

Abbott notes that some of these assumptions can be relaxed---for example, a time series allows an entity to change over time---but this does not substantially change the formulation of the world (at a given point in time, an entity is still fixed, and there is still no notion of multiple interpretations).\footnote{Another relaxation is from Karthik \citet{dinakar2017}, who proposes the idea of \textit{lensing} to represent multiple interpretations; while observations are still fixed, the value of a latent variable that represents cluster assignments of observations can be different for different annotators.} 

As an example of this formulation leading to conceptual problems downstream in statistics and econometrics (and, in this case, in applications of those in law as well), Issa \citet{kohler2019} critiques the ``counterfactual'' view of race. This view defines discrimination by asking if a person would have been treated if everything was the same (i.e., \textit{ceteris paribus}) \textit{except} for their race. From a sociologically rigorous point of view, the idea that one could hold everything but somebody's race constant is incoherent: aspects of people are so deeply tied into how their racial identity is constructed, and the ``effects'' of race so suffused into so many other variables, that a hypothetical ``race-switched'' world is meaningless (also, within a Rubin causal framework, race is immutable and not manipulable hence cannot have a causal effect). We would need to ``backtrack'' and revise all aspects of the person to make the counterfactual meaningful, but then there would be no pure effect to estimate.\footnote{One solution is to change the causal effect from race itself to perceptions of race; while clever, such a shift does not account for differences in life trajectories. That is, ``audit studies do not measure the objective-isolated treatment effect of race and race alone because there is no such thing to measure'' \citep{kohler2019}. Actual engagement with discrimination would come through traditions of antisubordination and cultural reconstruction.}

Lily \citet{hu2019a} extends this to machine learning, noting that auditing for fairness based on posing counterfactual in terms of model inputs suffers the same conceptual incoherence. She argues that this framing reduces discrimination to irrationality (i.e., discrimination relies on ``irrelevant'' characteristics in decision-making), egregiously neglecting the ways in which racism is very rational for maintaining power structures that bring benefits to certain groups. This problem can somewhat be alleviated by recognizing the ``right'' kind of dependencies between race and other variables, and structuring graphical models along those terms by which analysis is done, but so long as the model allows for statements to be made based on a generative case of two people who are alike in all aspects but race, the fundamental problem remains \citep{hu2019b}. Elsewhere, \citet{hu2020} argue the same problem of conceptualizing what makes a group and the assumption of modularity in causal models apply to sex as well. This argument would similarly extend to other identities (sexual orientation, trans identity, disability, etc.) and intersections thereof. 

This ties also into a larger problem identified by Philip \citet{agre1997} of technical disciplines trying to read social science and humanities literature as descriptions of mechanisms (or potential model formulations). Doing so fails to understand their content and, in the above case (where race is recorded as a column of a data matrix and treated as a variable), can even be inconsistent with it. 

\subsection{Central tendency}
The central mechanism of all statistics and machine learning is the use of a central tendency (justified by reference to probability distributions, and the ``sufficient statistics'' of a given distribution). For continuous variables, this would be the mean or median, or for categorical variables, this would be the majority class. The central tendency might be of a subset of the range of values (e.q., quantile regression, or mixture models), but is still a central tendency within that region.\footnote{The canonical parameters of a uniform distribution could be characterized as an ``extremal'' tendency rather than a central one, but such a distribution could also be equivalently described by the central tendencies of its center and \textit{radius}.} Outlier-based tasks like anomaly detection that identify individual observations still do so by reference to a central tendency. And estimators for supervised learning tasks or used for statistical inference are formed from the \textit{conditional} central tendencies at various levels of features/covariates. 

Both statistics and machine learning are fundamentally unable to do anything with one observation of one entity; there either have to be multiple observations of that entity, and/or observations of multiple entities (along with an assumption or definition of what constitutes an ``entity,'' and which such entities are comparable in a distribution of some property). While there are interpretations of probability statements as applying to individuals rather than to groups \citep{dawid2017},\footnote{I thank Joshua Simons for pointing me to this reference.} estimates of probability are still only able to be calculated from multiple observations.

Specifically: \textit{predictions are of the central tendency}. Despite promises of ``personalized'' medicine or ``individualized'' risk assessments,\footnote{Specifically, Philip \citet{dawid2017} says that the foundational philosophical question of of ``individual\textit{ized} risk'' is a notion of ``individual risk.'' For an individual, Sam, a notion of individual risk requires, for a Frequentist interpretation of probability, the assumption that ``the chosen attributes capture `all relevant characteristics' of the individuals.'' A Personalist interpretation of probability (a variety of a Bayesian interpretation) would require an assumption of ``no relevant additional information about Sam (or any of the other individuals in the data), and can properly assume exchangeability---conditional on the limited information that is being taken into account.'' Dawid notes, ``Neither of these requirements is fully realistic.'' Again, thanks to Joshua Simons for this reference.} treatment or scores are not really done by the individual, but by making finer and finer bins (which is possible given more and more data).\footnote{This phrasing came \redact{Elena Esposito}, in a conversation. My thanks to her for this.} There are questions of justice in applying inferences about a group to any individual member of that group \citep{faigman2014}; and at the point where the bin contains only one person, statistics and machine learning are incapable of doing anything but restating the data. Similarly, \citet{katz2017} talks about the limits of image recognition, selecting famous historical photos (e.g., around segregation in the US) to make a point about how image recognition not being able to understand the context of those photos: ``The presumption is that mappings from images to labels are sufficient; that the `information' is there, and it's only a matter of finding the right model to decode it,'' but one-off information is not what ``Statistical pattern recognizers are suited for.'' Alternatively, correlates of one-off contextual information are simply not present in the loss landscape of the model, so no optimization approach over a feature space derived from pixels could ever recover them.\footnote{I thank \redact{Beau Sievers} for encouraging me to think about the loss landscape.}

Some sort of central tendency will invariably be the minimizer of a loss function, so thinking in terms of loss functions \textit{aggregated across multiple observations} leads inevitably to central tendency. But such aggregate loss is a commitment to efficiency, and of a certain notion of utilitarianism (where the utility to be maximized is something measurable). It, too, is neither natural nor inevitable, and perhaps is quite unnatural: when the idea of characterizing a population by its central tendency first arose, the idea was met with great skepticism and opposition \citep{donnelly2016,hacking1990}, and represented a large shift in basic ideas about the world \citep{daston1989}. 

Even if unnatural, central tendency was a highly effective solution to a problem of too much data. In the time of R. A. \citet{fisher1922}, data sets (in the dozens of observations) were already too massive to interpret manually. He wrote, 
\begin{quote}
``briefly, and in its most concrete form, the object of statistical methods is the reduction of data. A quantity of data, which usually by its mere bulk is incapable of entering the mind, is to be replaced by relatively few quantities which shall adequately represent the whole, or which, in other words, shall contain as much as possible, ideally the whole, of the relevant information contained in the original data.''
\end{quote}
But extending the notion of minimizing loss, or to summarize information via a central tendency, from experimental design contexts to tasks of social management and \textit{governance} was a slow process for which certain commercial interests had to put enormous effort to succeed \citep{bouk2015}. 

Using the central tendency to determine outcomes has consequences: namely, it invariably punishes outliers \citep{keyes2018,costanzachock2018}. Is it possible to treat people as individuals at scale? Todd \citet{rose2015} gives an example at least of an amelioration: he describes the US Air Force designing plane seats around the ``average pilot,'' but then discovering that these averages did not fit a single pilot (and the design leading to deaths). Instead, \textit{adjustable seats} were developed as a way to account for variability. 

Incorporating variability may often improve over using the mean or median, but is not something fundamentally different. Variance---defined as the second central moment of a probability density/mass function---is itself a central tendency, and therefore still describing something about a population rather than an individual.\footnote{We might try to address problems with variability by estimating skewness (the third central moment), and problems with skew may be solved by estimating kurtosis (the fourth central moment), and so on; but perhaps the only way to account for all ways in which individuals are different is to have as many moments as observations---in which case we have done nothing but restate the data. I thank \redact{Beau Sievers} for a discussion that led to this point.}

The problem may fundamentally be one of scale: perhaps scale inevitably requires standardization, and standardization is most effectively achieved through managing via central tendencies. Perhaps the only way to treat people as individuals requires giving up scale and centralization, and allowing systems and analysis to be local. This, however, is a very different picture of civilization. 

\subsection{Alternative forms of modeling}
Other forms of quantitative modeling may avoid these limitations, although they come with their own problems. There is a style of modeling of social systems, found in ``social physics,'' microeconomics, and mathematical sociology, that has been called ``mathematical'' \citep{kolaczyk2014}, although probability-based models are also mathematical so a better term would be helpful. This style consists of using linear algebra and/or calculus to express core assumptions (e.g., as equations) and then making derivations from them to reach conclusions. They potentially also use distributions to get a range of results, or use simulations to get plots of functional forms that cannot be calculated analytically, but in all cases, the focus is on the mathematical expression of processes and relationships. But this style also divides the world into entities and properties of those entities---just as mechanics, for example, has objects that have have properties like mass or momentum.\footnote{Even network models do not escape this reductionism; they move what counts as an entity from individual people to dyads, triads, or even whole networks, but they are still reductionist to those \citep{erikson2013}, and still do not allow for multiple interpretations or mutable entities.}

And, as compared with statistics, the arguments of this literature are not necessarily applicable to understanding any specific real-world system \citep{borgatti2009}. Indeed, it is a whole separate question of whether a given structural property and the kind of generating processes implied by it---such as ``power law distributions''---could even apply in a specific case, or indeed ever applies to meaningful processes in the social world \citep{clauset2009}. 

Simulation modeling, especially as realized in agent-based modeling or agent-based social simulation, is a distinct alternative to the formulations of probabilistic modeling.\footnote{Note that simulations use statistics, and statistics uses simulation, but as forms of \textit{modeling} they have very different goals and assumptions; see \citet{gilbert2005}} Here, modeling is of interactions between software agents, offering an alternative to relationships between variables \citep{macy2002}, whether done through statistics or expressed through equations. It is useful for proposing causal mechanisms, and can be useful precisely when experimentation is unethical or infeasible. However, it does not involve fitting to data, only initializing with data and then perhaps qualitatively comparing simulation distributions to observed ones; and it is nearly useless for specific, real-world prediction \citep{pfeffer2017}. It is therefore best being used as a method of \textit{theory development} \citep{gilbert2005}, and is not applicable for most cases where statistics and machine learning are used. To believe otherwise, and think that such an abstract simulation might give insights into a specific real-world system, is an unjustified leap of logic that risks dangerously distracting from processes at play in the world \citep{shalizi2011}, such as thinking that the Shelling model of segregation might describe segregation in Chicago rather than identifying redlining as the cause. Much earlier, sociologist Paul \citet{starr1994}\footnote{I thank \redact{Uri Wilensky} for this reference.} writes about the ``seductions of sim'': how the ``extraordinary variety and intricacy'' of simulations lead to the inevitable simplifications and assumptions, which only the technical creators properly appreciate, becoming naturalized in policy planning. Similarly, Ben \citet{green2019b} writes about how simulations of urban traffic flows are conspicuous in the simplification of excluding pedestrians---and how this encourages planning around cars, rather than people. Perhaps it is unfair to identify ``being too seductive for what it is actually based on'' as a limitation, since that seductiveness comes partially from being more accessible for non-modelers and thereby overcomes a ket limitation of other forms of modeling. But, if engagement comes from the rhetorical impact of the simulation rather than its expression and exploration of assumptions, it is not genuine engagement. A similar danger also appears around interpretability in machine learning, which I discuss below in section (\ref{sec:interpret}).

\section{Prediction and explanation}\label{sec:pred}
Today, statistics and machine learning have enormous, if not complete, overlap in their underlying machinery. While machine learning started off trying to understand cognitive processes of learning, from the 1980s on, a strand of machine learning that used statistical models \citep{boden2006} in (arguably) instrumentalist ways \citep{jones2018} to mimic aspects of ``learning''---namely, improving with ``experience'' \citep{mitchell1997}, which became operationalized as ``data''---proved successful at many target tasks that had otherwise eluded artificial intelligence researchers \citep{norvig2010}. This statistical version of machine learning became the overwhelmingly dominant form of machine learning and, indeed, AI.\footnote{While other versions and visions of machine learning and AI exist, I do not discuss them here.} As part of this shift, machine learning adopted advanced statistical concepts in the span of only a few years \citep{wasserman2014}, completing the convergence. The result is that the use of models like linear regression, or logistic regression, could either be statistics or it could be machine learning, depending on \textit{how} it is used, and for what purpose.\footnote{Confusion around this is shown in a recent exchange of \citet{christodoulou2019}, \citet{bian2019}, and \citet{vancalster2019}. I thank \redact{Muhammad ``Tuan'' Amith} for these references. As \citet{bian2019} rightfully point out, just because something uses logistic regression does not make it statistics (especially if it includes regularization, which is more appropriate for the goals of machine learning). The difference, to use the articulation of \citet{mullainathan2017}, is in which part of the model matters; if a fitted linear regression is $\widehat{\y} = \X \widehat{\bbeta}$, then statistics cares about the $\widehat{\bbeta}$ (which captures relationships between inputs and outputs) representing the relationships present within the system, whereas machine learning cares about $\widehat{\y}$ (the match to the target system). That is: even though the model class is the same, because of the differing goals, the final models will likely have different terms and weights. And it is surprising that there is a trade-off between these goals, but (partially based on definitions of success; see below) there is \citep{shmueli2010}. I thank \redact{Doaa Abu Elyounes} for noting the lack of clarity on this point previously. There are other stylistic differences as well, such as statistics focusing on quantifying uncertainty, and machine learning paying heed to efficiency in implementation.} Also as a consequence, the ``learning'' of machine learning is now a metaphor for optimization, bearing almost no resemblance to what is studied under socio-cultural learning theory. And unlike the goals of earlier machine learning, it arguably has vanishing relevance even for what is studied under cognitive learning theory \citep{marcus2018} beyond the level of loose analogy \citep{watson2019}. 

The goal of statistics, traditionally, has been to use data to understand something about underlying processes \citep{fisher1922}. For this, quantifying uncertainty has been central; how well a model generalizes to new data (or, mimicking having new data by use of a held-out set) can be a good argument in support of the model's validity, but it makes perfect sense to, say, model data within a unique system \citep{kass2011}. In contrast, while using much of the mathematical machinery and statistical models, machine learning seeks to design reliable input-output mappings, what Leo \citet{breiman2001} called ``algorithmic modeling,'' and the emergence of which represented a distinct historical development \citep{jones2018}.\footnote{Nonparametric statistics is frequently much more like machine learning, in often being much more useful for input-output mappings than for understanding underlying processes (although it is still distinguished from machine learning in usually including a quantification of uncertainty). Consider even a bivariate nonparametric regression, applied to data generated from a sinusoidal process: it would fit a curve close to the sinusoidal functional form automatically, but the model would not tell us in substantive terms that the functional form is, indeed, sinusoidal. Consequently, the fit would be mainly useful for prediction, and not insight into the underlying process. In this simple case, the data and fit could be plotted to perhaps visually identify the specific functional form (which is powerful if the data cannot be modeled as coming from a simple, known functional form), but this becomes harder and harder in higher dimensions. There, nonparametric models really only become useful for prediction, or exploratory analysis.} Most machine learning research is ``supervised learning,'' which is only meaningful if the systems or procedures demonstrated were actually to be deployed in an out-of-sample setting. Without demonstrations of success in applications settings, machine learning predictions ``are not prediction at all!'' but simply post-hoc claims that a prediction ``could have been made'' \citep{gayoavello2012}.

Machine learning having a narrow standard of success does provide many benefits: it means machine learning can arguably ignore the many problems that have been the focus of criticism within statistics, such as problems with quantifying uncertainty \citep{mallows1998,freedman2009}, interpreting probability \citep{freedman2009,cox1990}, and determining what the ``inferences'' of statistical inference are to \citep{kass2011}, especially given model misspecification \citep{buja2014}; so long as out-of-sample success can be demonstrated, none of that matters. The same applies for different types of validity; only external validity (generalizability) matters \citep{domingos2012}. Even face validity can be ignored. If we successfully showed that astrology (or rather, a machine learning model fitting astronomical data to real-world outcomes) had robust out-of-sample success, then for machine learning purposes, it would be perfectly valid to use. 

While both machine learning and statistics hold that there is an external world that takes primacy \citep[i.e., over actors' interpretations and renegotiations of it; ][]{payne2004}, they differ in how they treat this. Also distancing machine learning from statistical critique is how statistics---including the mainstream of Bayesian statistics \citep{gelman2012}---has a strongly realist bent, hypothesizing an underlying truth from which data are generated, and seeing the task of statistics as recovering this truth \citep{lavine2019}. The role of statistical theory is to understand how well a method recovers this underlying truth. In contrast, machine learning's internal definition of ``improving with experience'' \citep{mitchell1997}, and indeed the very metaphorical use of learning, is more instrumental \citep{jones2018} and even agnostic about truth; a machine learning model simply improves its performance with more data, and does not necessarily uncover some underlying truth (with the exception of the subfield of causal discovery and causal learning, and sometimes claims from unsupervised learning\footnote{Unsupervised learning, which corresponds to clustering within statistics, is usually not associated with claims of ``prediction.'' It might be associated with claims about uncovering underlying structure, but there is ample work \citep{vonluxburg2012} about the difficulty of validating clusters beyond the face validity of ``making sense.'' Sometimes, clustering might be used to find structure that then is used as an input into a predictive model (e.g., dimension reduction techniques); but in this case, the final task of prediction is the standard of success. Word embeddings, a representation derived from ``unsupervised'' learning, work fascinatingly well for many language-based prediction tasks; however, it is actually a supervised learning task of imputing a held-out word, and beyond its impressive successes, word embeddings have key limitations for what might be said to be actually \textit{understanding} language \citep{tenney2019,mccoy2019,glockner2018}. It is also possible to take an unsupervised approach to prediction (e.g., outlier-based anomaly detection for fraud): in those cases, a ``ground truth'' set of labels exists (or can be determined, such as by manual investigation) against which cluster assignments can be validated. Then, claims of success are again post-hoc statements about success that could have been achieved if the system were applied. I thank \redact{Doaa Abu Elyounes} for encouraging me to clarify this.}). Machine learning theory, then, is focused on what the maximum possible \textit{performance} is, and whether/how it might be achieved.

But the focus on prediction brings with it as many or more problems than it avoids. There are two issues. The first is that predictions, by virtue of being correlations, cannot provide explanations of the system, as most \citep[but not all; see][for a review]{woodward2003} philosophical treatments of ``explanation'' relate it to \textit{causality} \citep{alexandrova2009,reiss2012,alexandrova2013,reiss2013,miller2019}.\footnote{I thank \redact{Hemank Lamba} for the latter reference.} Defining causality is its own challenge, but effective for many uses is to do so in terms of manipulating, controlling, or changing things in the world \citep[again, see][]{woodward2003}.\footnote{I thank \redact{Lily Hu} for this reference. See also section (\ref{sec:cause}) below.} The second is that, even if our only goal is prediction, correlations are fragile basis for doing so. 

\subsection{Prediction is not explanation}
As Zachary \citet{lipton2018} notes, ``The optimization objective for most supervised learning models... is simply to minimize error, a feat that might be achieved in a purely correlative fashion.'' In fact, correlations are the \textit{only} way in which machine learning seeks to minimize error. Defining the output of optimized models as `prediction' is to define predictions as extrapolation (or perhaps interpolation) from correlations on a system assumed to be fixed, and defining prediction thusly is not universal. Indeed, in other disciplines (like physics, or economics) and perhaps in lay understanding as well, `prediction' is assumed to be referring to causality: consider how \citet{friedman1953} talks about prediction \textit{in the presence of change}, which is not the machine learning sense of the word. In physics, predictions are frequently of the outcomes of an experiment, or are around surprising \citep{salmon1981} or ``novel'' predictions \citep{alai2014}, like of the gravitational bending of light \textit{predicted} by general relativity \citep{brush1989}. That is: if predictions were defined differently, the conditions of success would be very different, and machine learning would look very different \citep{dotan2020} or perhaps not be distinct from statistics at all. 

Consider a case discussed by \citet{pena2019}, where an anti-abortion conservative Argentinian politician hailed a system using public databases and Microsoft Azure system to predict teenage pregnancies (as part of his opposition to a proposed law legalizing abortion). Of the system, the politician said, ``With technology, based on name, surname and address, you can predict five or six years ahead which girl, or future teenager, is 86\% predestined to have a teenage pregnancy.''\footnote{In Spanish, ``Con la tecnolog\'{i}a vos pod\'{e}s prever cinco o seis a\~{n}os antes, con nombre, apellido y domicilio, cu\'{a}l es la ni\~{n}a que est\'{a} un 86\% predestinada a tener un embarazo adolescente.'' The Spanish \textit{predestinada} has the same metaphysical connotations as the English ``predestined.'' I thank Joana Varon for confirming this, and for pointing me to the original quote.} This was in support of a frequently advanced anti-abortion activist position that \citet{pena2019} articulate as: ``if [governments] have enough information from poor families, conservative public policies can be deployed to predict and avoid abortions by poor women.''

It is striking to see prediction explicitly equated with \textit{predestination}, which is far beyond what machine learning predictions actually mean or do. And, it is also striking to note that this false certainty about ``predictions'' allowed machine learning to be rhetorically deployed in support of a regressive political agenda.

Talking about ``prediction'' can obscure how it is based on correlations. After all, how could a `wrong' model predict well? Even further, we can use the bias-variance tradeoff to explicitly derive `false' (misspecified) models that predict better than `true' models \citep[i.e., the specification of the data-generating process fit back to the data it generated;][]{shmueli2010}. This is true even asymptotically, although (if the estimator is consistent, and within the same model class as the true model) the gap in performance becomes negligible. Practically, though, for finite-sample prediction (as all prediction ultimately is) there is a ``conflict between the aims of optimizing finite-sample prediction and of finding the true data-generating model class'' \citep{kunst2008} of a magnitude large enough to matter. Elsewhere, \citet{mullainathan2017} discuss the division as ``$\widehat{y}$'' problems and ``$\widehat{\beta}$'' problems, noting that very different models, with very different causal implications (and hence, implications for public policy interventions) can achieve minimum loss equally well \citep[see also what][terms the ``Rashomon effect'']{breiman2001}. 

Philosophically, Nicholas \citet{rescher1998} theorizes ``trend projection'' and ``curve fitting'' \citep[which][critique mainstream machine learning as being limited to]{pearl2018} as \textit{rudimentary} or \textit{elementary} types of prediction, ranking underneath the \textit{scientific} or \textit{sophisticated} types of prediction of ``indicator coordination'' (with a linking mechanism of \textit{causal} correlations), ``law derivation'' (with a linking mechanism of deterministic or statistical laws), or ``phenomenological modeling'' (with a linking mechanism of formal mathematical/physical models). We can take or leave the normative implication of this progression, but either way, Rescher's levels are a useful way to relate what predictions mean in machine learning from they mean in physics or classical economics. 

For readability, in the remainder of the paper, I will be using ``prediction'' in the technical sense of the output of a correlation-based model (i.e., a ``rudimentary'' type of prediction) without scare-quotes.\footnote{I thank \redact{Sabelo Mhlambi} for pointing out that I do this, and the need for clarification.} But the reader should fill in such scare quotes for the remainder of this paper and, indeed, in any place the word is used in relation to statistics and machine learning. 

While it may seem like we can gain substantive knowledge about a phenomenon or entity based on which feature sets give superior predictive performance, the definition of prediction essentially makes this into interpreting causation from correlation. It is a worthwhile \textit{exploratory} exercise to hypothesize what causal mechanisms might generated observed correlations, but it is only that, and should not be presented as definitive \citep{lipton2018}. Machine learning papers with titles like ``Understanding...'' or ``Explaining...'' or even ``Studying...'' or ``Characterizing...'' are thus potentially misrepresenting or misunderstanding what can be concluded from the methods used. 

\subsubsection{Interpretable/Explainable is not the same as explanatory}\label{sec:interpret}
Furthermore, separating out the explainability or interpretability of a \textit{model} from the idea that the model explains the underlying \textit{system} is a difficult task.\footnote{Following most treatments, I do not commit to a specific definition of interpretability or explainability \citep{doshivelez2017,rudin2019}, although I take fitted decision trees (as in the quote from Breiman) or decision rules derived from such trees \citep[as in][]{letham2015} as exemplars of being inherently ``interpretable,'' or equivalently, of providing their own ``explanations.'' I will note that causality is usually quoted as one potential criteria for interpretability (in which case decision trees/rules would \textit{not} be inherently ``interpretable''!), but causality (or even an attempt thereof) is a much more difficult task than any other notion of interpretability. So, it would be surprising if a model that meets this higher threshold were presented in terms of the lower threshold. Conversely, insofar as causality might be defined in terms of manipulations and interventions, and those are defined in terms of potential human actions, any causal model will likely be ``interpretable'' under other notions of the term (at least in its parts, if the whole is, for example, not sparse).} Leo \citet{breiman2001}, in hailing the interpretability of decision trees, also provides an example of this breaking down:
\begin{quote}
``A project I worked on in the late 1970s was the analysis of delay in criminal cases in state court systems... The dependent variable for each criminal case was the time from arraignment to the time of sentencing. All of the other information in the trial history were the predictor variables. A large decision tree was grown, and I showed it on an overhead and explained it to the assembled Colorado judges. One of the splits was on District N which had a larger delay time than the other districts. I refrained from commenting on this. But as I walked out I heard one judge say to another, `I knew those guys in District N were dragging their feet.'\thinspace''
\end{quote}
Breiman does not remark further on this, but I would say: was District N actually dragging its feet, i.e. were faster criminal trials within their ability but simply not done? To properly answer this question would be to provide an estimate of the effect of being in District N, after controlling for other factors (such as, say, amount of resources). But a decision tree might use an indicator for being in District N as a proxy for the actual causal factors (with which being in District N correlates highly) and thereby achieve the same final predicted values. In this story, the judge interpreting the explainable decision tree arrived at an unjustified conclusion. That is, maybe the conclusion was actually accurate, but using the decision tree was not a valid way to make that determination.\footnote{Consider also \citet{letham2015}, who present interpretable decision rules for classification using the \textit{Titanic} dataset as an example. They write, ``For example, we predict that a passenger is less likely to survive than not \textit{because} he or she was in the 3rd class'' [emphasis original]. The ``because'' refers here to the \textit{model's} behavior, not to causality (which the model is not attempting to capture), although in this simple case they likely coincide.} This also relates to notions of model ``correctness'': as a description, a bivariate correlation is simply an observation and (within its way of framing the world, and dependent on the meaningfulness of the data) is ``correct,'' but this examples shows that this notion of correctness is deeply counter-intuitive, and, as I discuss below around interpretability, perhaps too narrow and fragile to be an acceptable standard. 

This is a running, but backgrounded, tension in the interpretability/explanability literature. \citet{doshivelez2017} write, ``one can provide a feasible explanation that fails to correspond to a causal structure, exposing a potential concern.'' \citet{caruana2015} write, ``Because the models in this paper are intelligible, it is tempting to interpret them causally. Although the models accurately explain the predictions they make, they are still based on correlation.'' And \citet{lipton2015} writes, ``Another problem is that such an interpretation might explain the behavior of the model but not give deep insight into the causal associations in the underlying data... The real goal may be to discover potentially causal associations that can guide interventions.''

\textit{If} there is a problem for which causality is not the end goal, then \citet{rudin2019} provides a good argument for ``inherently interpretable'' models (like decision rules) over explanations of black box models---and in so doing, daringly challenges the premise and very foundation of an enormous emerging area of research around ``explainable AI.''\footnote{For \citet{rudin2019}, \textit{interpretable} models are ones that ``provide their own explanations, which are faithful to what the model actually computes'' (which would certainly agree with fitted decision trees, but not necessarily with some of the other provided potential criteria she quotes, particularly causality). Whereas an \textit{explainable} model is when a second model can be created to explain a first model. In this sense, an explanation of a model (i.e., the second model) will almost by definition not faithfully convey what the (first) model actually computes (whether through necessity, or because of obfuscation), otherwise there would be no need for the second model! Also, note that the sense of ``explainable'' in the subtitle of this section, and the exemplar of fitted decision trees, is closer to Rudin's sense of ``interpretable.''} However, while accepting the superiority of interpretability of explainability as she defines them, I would add a similar challenge to interpretability as well: \textit{interpretability is irrelevant if the audience does not understand the distinction between correlation-based predictive models, and models that try to capture causality or something about the data-generating process}. Indeed, because of correlations between variables in data, many ``interpretable'' models with very different sets of variables and weights on them will provide similar predictive performance, and yet none may correspond to a causal process. I agree with \citet{hancoxli2020}  in that the ``moral hazard'' of being able to select a model that looks like it is using ``acceptable'' covariates, e.g. postal code instead of race (for which postal code can be an almost perfect proxy), over a model that more accurately reflects the underlying (causal) relationships being leveraged, is again a danger of interpretability not being causality. As in the story of Breiman above, if a predictive model is interpretable, it provides the danger of the \textit{illusion} of engagement. This illusion gives similar end result to the danger, as discussed by Rudin, of black box models being pushed through on the basis of authority/intimidation: that of denying stakeholders any real opportunity to engage. 

``Interpretable'' models potentially have a redeeming feature around causality: because they can be more easily critiqued by non-modeling experts, if an objection to a model arises based on the model violating causal intuition or domain knowledge, it would be an opportunity to explain to stakeholders that the model is based only on correlation.\footnote{I thank \redact{Cynthia Rudin} for an exchange that led to this point.} But if the distinction could \textit{not} be successfully conveyed (e.g., the audience kept insisting on taking interpretations as being of causal relationships in the world), or if the audience rejects the model once they understand it is based only on correlations, it would show the modeler that a non-causal model is not good enough for the application. Again referring to the Breiman story, if the audience does not already understand that what ``predicts'' is not what causes (or, as above, what is explainable is not explanatory), or the modeler does not successfully explain this, either explainability or interpretability are a dangerous distraction, too narrow in its goals and fragile in how it is understood.

\subsection{Correlations can fail}

\subsubsection{Fragility of non-causal correlations}
Non-causal correlations are traditionally known as ``spurious correlations,'' i.e. a correlation that empirically exists and is robust not because of a direct causal link, but by some structure like a common cause \citep{pearl2009}. Murders and ice cream sales may indeed correlate, but the underlying cause is hot weather. Per capita, Nobel prize wins and chocolate consumption are correlated \citep{messerli2012}, but (despite Messerli's joking post-hoc explanation about certain chocolate proteins stimulating brain function) the underlying cause is likely access to resources. However, ``spurious correlations'' are also used to refer to correlations that are the result of improper data mining, i.e. that are \textit{not} robust do not generalize \citep{fan2016}.\footnote{While the ``robust but not causal'' sense of spuriousness is more common, I note that the famous article of Udny \citet{yule1926} talks about ``nonsense-correlations'' in the other sense: ``if we had or could have experience of the two variables over a very much longer period of time we would not find any appreciable connection between them.'' This is also the sense used by the book of \citet{vigen2015}, although the book does not get into why these correlations are spurious. Yule's explanations were ultimately getting at what we would today call temporal autocorrelation (and cyclic behavior); two time series can appear correlated with each other when in fact they only have similar autocorrelation or cyclic structure. One solution to this is to think in terms of effective sample size; if we calculated confidence intervals for correlations that took the effective sample size into account, the correlations may be large but still not significant, even before including a correction for multiple comparisons.} For clarity, I will refer to these robust correlations as ``non-causal'' rather than spurious. 

The danger of using non-causal correlations alone for prediction, as is also a central argument in causal learning literature \citep{spirtes2016}, is that predictions will not be robust to changes in context and over time. In such cases, underlying causal processes will lead to ``distribution shifts.'' Bernhard \citet{scholkopf2015} references old behavioralist language \citep{katz2017} in analogizing this to ``superstitious behavior,'' ``in which statistical associations may be misinterpreted as causal'' \citep[although I would say that just because a model \textit{uses} correlation for prediction, it does not mean we should speak anthropomorphically of the model as ``interpreting'' the correlation as causal;][]{watson2019}. And Jonathan \citet{zittrain2019} proposes the idea of ``intellectual debt'' to describe the dangers of coming to rely on systems that work through correlations whose underlying causal structure we don't understand. 

An excellent example of this is in the case of Google Flu Trends. Here, the cross-validation was done correctly (as I discuss in section \ref{sec:CV}), in temporal blocks \citep{ginsberg2009}. Nevertheless, the system failed when deployed: as it turned out, the model was relying on a strong correlation between winter-related features and incidence of flu, not having observed the variability represented in cases where these two did not correlate. \citet{lazer2014} phrase this as a problem of overfitting, and indeed we can think of models that fit to anything other than an underlying causal process as ``overfitting,'' but it is more strongly a lesson about causality \citep{harford2014}. As more data are gathered, capturing more of the total range of variability, the system would likely improve; but there might be some other causal processes because of which the system fails later on. 

\subsubsection{Correlations are proxies and can be gamed}\label{sec:gaming}
Another failure point is in how correlations, like any proxy, can be gamed. And, if there are any stakes associated with a proxy, it is nearly a guarantee that they \textit{will} be gamed in a market system. ``Goodhart's law''\footnote{Originally, ``Any observed statistical regularity will tend to collapse once pressure is placed upon it for control purposes'' \citep{goodhart1974}.} is that ``When a measure becomes a target, it ceases to be a good measure'' \citep{strathern1997}. This relates also to earlier work by \citet{ravetz1971}, and to the ``corrupting effect of quantitative indicators,'' about which \citet{campbell1975} wrote (emphasis original), ``\textit{The more any quantitative social indicator is used for social decision-making, the more subject it will be to corruption pressures, and the more apt it will be to distort and corrupt the social processes it is intended to monitor}.''

Take the example from above (fig. \ref{fig:cat}) of image recognition: neural net models find groupings of pixels that correlate with laboriously collected human-given labels \citep[labor that][refer to as ``automation's last mile'']{gray2019}. The groupings of pixels are only a proxy for the image content and neither the content itself nor analogous to human perception; as a consequence, adversarial examples can exploit this \citep{goodfellow2015,eykholt2018} and add noise imperceptible to human eyes but that can cause an image (or, indeed, a 3D-printed object) to be arbitrarily misclassified as anything an attacker wants. 

Indeed, one reason for keeping credit scoring systems secret is because knowing their operations would make it easy to game them by optimizing to the proxies that they ultimately use, which is far easier than optimizing the underlying construct (not so much creditworthiness---which implies a moral judgement and should probably be done subjectively---as whether or not somebody will repay a loan, which is ``objective'' but which only time can reveal). Cynthia \citet{rudin2019} writes, ``The reason a system may be gamed is because it most likely was not designed properly in the first place, leading to a form of Goodhart's law if it were revealed... transparency could help improve the quality of the system, whereby attempting to game it would genuinely align with the overall goal of improvement. For example, improving one's credit score should actually correspond to an improvement in creditworthiness.'' But that does not address the issue of \textit{constructs} (or causality). Creditworthiness is a construct that does not have a physical reality on which improvement can be known, so we will always need proxies, and therefore will always suffer Goodhart's law. Measurements of the physical quantity of ``whether or not a loan a given loan will be repaid,'' which is what a measure of creditworthiness is supposedly useful for predicting, are what are used for training models (or rather, past instances of such repayments). Yes, for potential borrowers it would be ideal that they only be judged on the actions they take that make loan repayment more likely; but actual repayment is likely heavily influenced by factors outside of a potential borrower's control, such that what minimizes loss for lenders is to use circumstantial information. This means that lenders will prefer using non-causal, potentially black-box correlations that they will keep secret, rather than trying to develop a ``properly designed system'' by the above criteria. In terms of causality as well, if we want to be able to know how to effectively make an \textit{intervention} into the system (to raise a credit score), then we would want a causal model, not one based on correlations, as I discuss below.

\subsection{Correlations are not a reliable basis for intervention}
Under this notion of prediction, another consideration of what predicts not corresponding to causal processes is that predictions are unable to guide \textit{interventions}. \citet{barabas2018} discuss how risk assessments in the criminal justice system are also used for \textit{diagnostic} purposes and in deciding how the criminal justice system intervenes in the lives of offenders; but this is unreliable, because risk assessments are based only on correlations. \citet{mullainathan2017} give an example of applying the lasso to random subsets of the American Housing Survey data. In each subset, a very different set of variables are selected in, all giving equally good predictive performance\footnote{I.e., a case where there is no ``stable set'' \citep{meinshausen2010}.} but, as they point out, giving very different implications for intervening. Indeed, the lasso (and any variable selection method) has no guarantees that it will select causal variables (if they are even among the measurable and measured variables); consistency guarantees are to the \textit{oracle} predictor, the model giving the best possible (post-hoc) performance within the given model class \citep{zhao2006}. As above \citep{shmueli2010}, the ``true'' model is not necessarily the best-predicting model, asymptotically and certainly in finite-sample prediction \citep{kunst2008} even if the estimator is consistent and in the correct model class. Of course, all bets are off if the estimator is not of the true model class, if the causal variables are not measurable or not measured or, more philosophically, if the true data-generating process cannot be described by \textit{any} functional form, let alone one that is smooth enough to reliably estimate from available amounts of data.

\subsection{What about size?}
What is the relationship of the size of available data to these problems? Not much; size has little to do with causality. In statistical terms, the main benefit from size is that it allows inference for characteristics of smaller and smaller subpopulations \citep{cox2015}. For machine learning, it allows greater and greater accuracy for predictions on such subpopulations. 

Causality aside, size tending to the entire population does not even overcome selection bias, as Xiao-Li \citet{meng2018} shows in recent work. He breaks down the estimation error (the difference between an estimate and a true underlying population parameter) as product of three quantities: the correlation between whether people respond and what their response is (a ``quality'' index, e.g. people who vote for an unpopular candidate are less likely to report their choice honestly), the square root of the difference between the sample size and population size (a ``quantity'' index), and the standard deviation of the underlying quantity (a ``difficulty'' index). With this, he shows that there are three ways to make the expected error be zero: have the entire population, have a homogenous population (the standard deviation is zero), or have no correlation between whether people respond and what their response is. Conversely, while randomization guarantees the correlation between responding and response is zero, this term is otherwise $O(1)$. It will never go to zero, even in bigger and bigger populations. 

\subsection{Prediction policy problems?}
Are there cases where correlation alone is sufficient? Alongside the obviously wrong claim in the 2008 ``End of Theory'' that with enough data we don't need ``models,''\footnote{Any use of data requires summarizing or reducing, e.g. into a prediction. And any summary of data is necessarily a model \citep{fisher1922}. ``End of Theory'' also starts with the \citet{box1979} quote, but completely misunderstands what a model is, or the relationships of models to data.} is the claim that ``correlation is enough.'' There is a legitimate version of this argument: \textit{sometimes}, under specific conditions, correlations \textit{may} be sufficient (although, again, this has nothing to do with the size of the data). \citet{breiman2001} argued this, writing about ``\textit{prediction problems},'' where focus on trying to approximate a data-generating process was unhelpful for the application; problems like detecting toxicity of chemicals in water, or speech recognition. He argued that statisticians have neglected these problems, and development of techniques for these settings ``has occurred largely outside statistics in a new community---often called machine learning---that is mostly young computer scientists.''

While not citing Breiman, \citet{kleinberg2015} make a nearly identical argument fourteen years later, and to an audience of econometrics and public policy rather than statistics: that ``there are also many policy applications where causal inference is not central, or even necessary... Not only are these prediction problems neglected [in public policy], machine learning can help us solve them more effectively.'' And, similarly to how Breiman argues that predictive solutions can nevertheless provide ``information'' (which I read as exploratory insight) about the underlying system even if that is not necessary or an explicit goal, \citet{kleinberg2015} argue that the solutions to prediction policy problems can provide ``not just policy impact but also theoretical and economic insights.''\footnote{Similar papers, arguing for the neglected importance of prediction, are recently appearing for many other fields as well such as \citet{yarkoni2017} for psychology and \citet{bzdok2018} for biology. Of these two, only the former is aware of and cites Breiman as precedent for this argument, but both conflate prediction with understanding, failing to acknowledge how a model can ``predict'' well without corresponding to a causal process.}

They systematize the conditions for such ``prediction policy problems,'' metaphorically labeled ``umbrella problems'' (in contrast to causal problems, metaphorically labeled---in move that is unfortunately more primitivist than it is humble about causal inference---``rain dance'' problems): our intervention must have no effect on the outcome, such as how umbrellas have no effect on rain. 

Are there really such problems? \citet{breiman2001} gives ``sources of delay in criminal trials'' as an example of a ``prediction problem,'' but ``sources'' is a causal (``$\widehat{\beta}$'') question and as I argue above, what the clients (the judges) really wanted was causal knowledge! Correlations might still be \textit{useful} if used as exploratory, but they would not be sufficient for guiding interventions. Notably, while he gives other candidate tasks that are promising, they are of physical and not social systems (using mass spectra, radar, sonar).\footnote{Another example he gives, ``On-line prediction of the cause of a freeway traffic breakdown,'' has the word ``cause'' but presumably the cause of a given breakdown would be a human-coded label, which would be what a model would try to recover as a mutliclass problem, rather than something inferred from data that measures aspects of a traffic breakdown.} And, as referenced earlier, \citet{barabas2018} discuss how, in practice, risk assessments in criminal justice are used for deciding on interventions; if this is how they are being used, risk assessment is not a prediction policy problem. 

\citet{kleinberg2015} propose five examples that are also questionable. A full reconsideration of these examples is beyond the scope of this paper, but I mention some problems with a few of them. 

Their first example is ``predicting which teacher will have the greatest value added'' in education, which I would see as a causal question (what \textit{causes} teachers to add value?), but would also point out that the study they cite measures ``value'' by students' improvement on standardized test scores. This is an unreliable measure of the underlying construct of ``having learned'' \citep{jacobs2019,oneil2016}, and the uncertainty of this carries through to the potential prediction policy problem. 

Their third example is ``targeting health inspections'' in restaurants using Yelp review data,\footnote{While the paper they cite has recently been disproved in a re-analysis by \citet{altenburger2019}, who show reviews provide no stronger a correlation than restaurant characteristics, those restaurant characteristics could still be the basis for this being a prediction policy problem.} but this setup is (potentially) adversarial, with any proxy being susceptible to gaming and manipulation. To return to the ``umbrella problems'' metaphor, while clouds don't determine whether or not they rain based on the presence of umbrellas, restaurants could make compliance decisions based on how inspections are directed. 

Their fifth example is ``lenders identifying the underlying credit-worthiness of potential borrowers.'' But this is both about an underlying construct (credit-worthiness), and a construct for which the historically and currently used proxies have caused and continue to cause oppression and increasing inequality \citep{lauer2017,poon2007,fourcade2013}, which I further discuss below. 

Below, in section (\ref{sec:exp}), I give what I think is a better candidate for a prediction policy problem, where the data and system are sufficient to apply correlation-only models to questions of public policy and the social world: deciding whether or not to assign potential cancer patients to chemotherapy based on observed correlations between gene signatures and developing breast cancer. But, important to note, it took experimental validation (rather than the post-hoc back-testing of cross validation) to know how to properly use model outputs (only as a second pass after a clinical diagnosis, rather than exclusively or as a first pass), suggesting that even if a problem is a prediction policy one, it might still take experimental testing to be able to effectively use it. And, in some cases, if we need to do an experiment anyway, we might want to use that opportunity to instead estimate the causal effect of specific factor on which we can then intervene.

\subsubsection{Injustice of correlations}
Conversely, what happens when we apply a correlation-only approach to problems that are \textit{not} prediction policy ones? Predictions have consequences: as \citet{mcquillan2018} writes, ``The predictive nature of machine learning promotes preemption, that is, action that attempts to anticipate or prevent
the predicted outcome,'' which \citet{pasquinelli2019} interprets as transforming correlations into a ``political apparatus of preemption.''

For a case discussion, I take the last example of \citet{kleinberg2015} of creditworthiness. In recent work, Rodrigo \citet{ochigame2020} connects the history of credit to machine learning, particularly around ``actuarial fairness,'' and here I summarize and expand his arguments. Note that risk assessment in criminal justice has very much the same story, with many historical developments happening parallel to (and not independent of) those of credit; this connection is summarized by \citet{barabas2018}, who also show how the ethical issues and sociological consequences of risk assessment are instructive for how to think about machine learning. 

For credit, being given a loan is not decided by things like people's intention or effort towards repayment, which are the only things over which people have control; instead they are decided by things over which people potentially have no control, namely the central tendency of previous repayment behavior by those who share attributes with the individual. For the US, Josh \citet{lauer2017} details how creditworthiness decisions in the 1960s were made on the basis of judgements of ``character''; this was far from fair, but there could be notions of (and methods of making determinations about) character that would be a fair standard against which to hold individuals accountable. This approach, though, was gradually replaced by one based on optimization and correlation that was far more empirically successful, just as in risk assessment at the same time \citep{hannahmoffat2013}. 

Having such data-driven decisions, based on correlations that become how lending ``risk'' is estimated, is certainly more efficient and effective for lenders; but it effectively rewards and punishes people, shaping what \citet{fourcade2013} call ``life chances,'' for things over which they potentially have no control.

While the shift in the insurance industry happened in the 1960s, attempts at empirical justifications for differential treatment goes back further. Dan \citet{bouk2015} details how, in the 1880s and 1890s, insurance companies were charging African Americans more for life insurance (often without their knowledge), which was noticed by Massachusetts state representative Julius C. Chappelle, ``an African American born in antebellum South Carolina and a janitor for the state of Massachusetts by trade,'' who proposed a bill to ban the practice. 

\begin{quote}
``Frederick Homer Williams, a lawyer from Brookline, chaired the committee that opposed the bill... Williams cited statistics from around the nation showing shorter life spans for blacks, including 1870 census figures showing a 17.28 death rate for `colored people' against 14.74 for whites. These numbers, Williams argued, and not any `discrimination on the ground of color' motivated insurers' rates. It was a `matter of business,' and any interference, he warned ominously and presciently, `would probably cut off insurance entirely from the colored race.'

``Chappelle's allies noted that Williams's statistics, while bleak enough, answered the wrong question. The question was not whether blacks in slavery or adjusting to freedom were poor insurance risks, or even whether southern blacks were poor risks. The question was African Americans' \textit{potential} for equality and specifically the present and future state of Massachusetts' African Americans---about whom no statistics had been offered by either side.'' 
\end{quote}

While the mortality statistics were perhaps \textit{empirically} true, as a basis for justifying charging more, it exacerbated inequality and was therefore rejected. Here, the argument for equality won the day, with the bill passing and inspiring similar bills in other states. 

The move from a moral basis to an empirical basis that happened in the 1960s provoked similar opposition. In an argument summarized by \citet{ochigame2020}, Caley \citet{horan2011} details how, in the 1970s and 1980s, civil rights and feminist campaigners charged the US private insurance industry with discrimination; the industry responded by promoting the idea of ``actuarial fairness,'' which \citet{ochigame2018} summarize as ``the antiredistributive principle that `each person should pay for his own risk,' in opposition to the principle of solidarity and mutual aid \citep{stone1993}.'' \citet{horan2011} writes, ``Admitting that fairness might mean one thing socially and another `actuarially' allowed pragmatically minded activists a way out of sticky philosophical discussions and even stickier debates with insurers over highly technical aspects of industry. Yet this concession, at the same time, had a crippling effect on attempts at regulation. Once insurers were granted their own unique definition and claim to fairness separate from that of society writ large, the application of social legislation became nearly impossible.'' Continuing with their summary of \citet{horan2011}, \citet{ochigame2018} write, ``The industry not only... defeated the efforts of civil rights and feminist activists [and] evaded legal regulation but also cemented an understanding of private insurance as essential, natural, and detached from questions of social justice.''

In another facet of this debate, Barbara \citet{kiviat2019} details resistance from consumers and local lawmakers against the auto insurance industry adopting certain measures, namely credit scores; these indeed \textit{correlated} with previous higher insurance payouts, but seemed to industry outsiders to lack a causal connection to being a good driver, or be something to which it was fair to hold drivers morally accountable (and therefore fairly rewarded with lower insurance premiums or punished with higher premiums). And Devin \citet{fergus2013} details how US auto insurance is based more on postal code than driving record, which in effect is a ``ghetto tax'' that punishes female, black, and Latinx drivers, those least able to afford a higher burden. Postal code (and indeed, ultimately, race itself) may indeed correlate more highly with costs to insurance companies than driving records, because structural factors are frequently more important for outcomes than factors under people's individual control.

Today, credit decisions being made on an expanded set of types of data that can be correlated with previous repayments \citep{noppper2019} are the continuation of this historical legacy, with equally troubling implications. And, just as the insurance and credit industry are required to provide ``explanations'' for their decisions---explanations that merely cite correlations that, as work such as Kiviat's show, are accepted on the basis of causal intuition among audiences (that the industry is happy to encourage)---\citet{barocas2020} draw a parallel to explanations in machine learning. These may be similarly misleading to consumers in terms of failing to map to actions that a consumer could have taken (i.e., causal factors, over which a consumer has agency/control) to change how they were treated. 

A legal and philosophical analysis is given by Deborah \citet{hellman2008} in her book, \textit{When is discrimination wrong?} \citep[in an argument later also presented to a machine learning audience;][]{hellman2018}. She develops a theory of when discrimination (i.e., drawing distinctions between people and treating them differently on that basis) is morally permissible, versus when it is morally impermissible. She identifies a trait being ``relevant'' or ``irrelevant'' as an insufficient standard, since ``relevant'' could be defined simply as correlated, and ``sex is likely correlated with work schedules and the costs of childbearing'' yet (in this case) such reasoning has been rejected as contributing to marginalization and overall undesirable inequality. Of course, such a standard is exactly what ``actuarial fairness'' institutionalized, as discussed above. Hellman's solution to when a distinction is impermissible is when it ``demeans,'' which she defines as treating another in a way that, in a given context, denies equal moral worth (regardless of intention, and regardless of whether or not an affected person feels demeaned, stigmatized, or harmed). If we were to adopt this standard, we could combine this argument with that of \citet{fourcade2013} that ties being considered morally worthy to being given positive life chances, and say: classifications that negatively affect the ability of people to exist, survive, thrive, and change in positive ways of their own choosing are morally wrong, regardless of their empirical accuracy. 

That is, much of the ``fairness'' literature in machine learning perhaps concedes too much in presuming that we will, or should, be using the correlation-maximizing models of machine learning, and privileging efficiency and the ability of, say, insurance companies to minimize uncertainty around payouts and thereby increase profits. For many of the applications currently being discussed, perhaps what would be actually ``fair'' is to collectivize risk, or even to deliberately redistribute resources to those whose structural position puts them at the highest ``risk.'' This would be a completely different optimization objective, or perhaps a political task like prison abolitionism \citep{benjamin2019,ochigame2018} that falls outside of any optimization framework. 

\subsection{Optimizing to the status quo}
\citet{carr2014} critiques big-data approaches as ultimately encouraging us to ``optimize the status quo'' rather than challenge it, channelling the point of Paulo \citet{freire1970} critiquing teaching to the status quo. If the only available data reflect heavy social biases, taking previous success as the optimization objective will only reproduce those biases \citep{noble2018}, especially in machine learning where we rely on what correlates most optimally. Perhaps if we identify possible underlying causal structures, we can try to see how we might divert things from their current course. And, outside of empirical research or anything we can extrapolate from existing data, we can imagine wholly different futures through fiction and social theory. Within cultural products, optimizing to previously commercially successful music or to film scripts \citep{nyt2013} prevents investment in new art and risks making cultural products stale. And, outliers are by definition observations that fall far outside of a central tendency, making it difficult to use them as exemplars on which to fit a machine learning model. 

\subsection{Problems in implementation}
This paper is devoted to articulating the implicit assumptions made when using machine learning, and the problems of these assumptions. But I want to acknowledge that problems of \textit{implementation} are an additional layer. 

First, there are issues with implementation in software \citep{merali2010}. Documented examples of mistakes in scientific results mostly have to do with data handling, prior to going into software that does analysis and model-building \citep{miller2006,karraker2015}, especially when Excel is involved \citep{panko1998,hutson2010,herndon2013}. But mistakes happen in analysis as well \citep{qiu2019}. There are concerns about using out-of-date software libraries; \citet{irpan2018} gives the anecdotal example of finding a warning message ``saying my NumPy version was too old,'' and being ``told that sufficiently old versions of NumPy led to incorrect physics calculations that broke some baseline algorithms. No one understood why this happened, and no one thought it was worth debugging (and rightfully so) [versus just updating NumPy].'' 

Difficulty in doing something correctly can be considered a limitation, and indeed the difficulties of reproducible code and sensitivity to initial conditions in machine learning \citep{tian2019} is an increasing source of doubt for the overall reliability of machine learning research \citep{hutson2018}. 

But manual or mechanical computation as an alternative to software implementation is simply not feasible for the models used by machine learning---and, at this point, even for the models used in statistics \citep{efron2016}. For standard models (e.g., KDE, GLM, SVM, random forests, etc.), there is a choice about whether to implement them from scratch or to use standard software packages. But if available, in general there is no drawback to relying on standard packages that are often extensively verified, field-tested, and optimized. For more specialized models or field-specific applications, rewriting model implementations where feasible may provide needed robustness checks \citep{neupane2019}, but this can be seen as a part of best practice rather than a decision or assumption with limitations. 

For deep learning models, there may be a decision to be made about how much computational budget to use on searching through the solution space, as this has a huge impact on the quality of results \citep{dodge2019} but better results come at the cost of potentially massive carbon emissions for the energy the computation requires \citep{strubell2019}. 

But the biggest point of implementation is not in software or even computational infrastructure; it is in institutions and power dynamics, for which nontechnical parts of interventions supersede the importance of technical parts \citep{katell2020}. Even if we had a perfect prediction policy problem, reliable software, and competent implementers with the best of intentions, the system might still go awry. \citet{eubanks2018} describes three cases of computer systems for optimizing and automating decisions, detections, and resource allocations. While one of these cases seemed to be driven by bad faith (organized around a desire to see as much denials of benefits as possible), the other two projects (one implementing a matching algorithm and the other a machine learning model) had system designers with genuinely good intentions. Yet despite these, she describes the dehumanization felt by people subject to the systems, and the myriad ways in which the systems were not clearly better than what had come before, and were in some ways worse. 

We also cannot assume that models outcomes alone determine how people are treated; the ``human-in-the-loop'' can present another layer. For example, work looking at how judges use recidivism prediction systems for decision support \citep{cowgill2018} shows that judges do not ignore the predictions (if they did, what would be the point?), but nor do they adopt them entirely, and this selective adoption is nonrandom. More specifically, \citet{stevenson2019}\footnote{I thank \redact{Maria de Arteaga} for this reference.} show in Virginia that judges followed risk assessments a bit less than half the time. Unlike the program's stated goals, on average, the decision support system did not change the rate of incarceration, nor the length of sentences, nor rate of reoffends after release, just shifting who experienced these. High-risk people received longer sentences, and low-risk people received shorter sentences, with the two offsetting each other. One systematic deviation was in judges giving more lenient to younger defendants, despite youth having the highest correlation with reoffending; but as the study authors point out elsewhere \citep{vandam2019}, rationales ``like mercy for the vulnerable or evaluating the level of culpability that individuals have [based on maturity]'' suggest not treating young people more harshly than older people, \textit{regardless of the empirical reality around reoffending}. Adopting risk assessments more completely would have led to greater numbers of younger people incarcerated, and incarcerated for longer. 

However, vulnerability from systematic racism did not merit similar mercy. Even though the Virginia system did not use race as an input, after adopting risk assessments, black defendants were more likely to be incarcerated and with 17\% longer sentences (among judges who most adopted risk assessments). This also agrees with MTurk testing that shows higher predicted risks overriding priors more often for black defendants \citep{green2019}. Frank Pasquale \citep[quoted in][]{vandam2019} noted that ``algorithms end up being a rationalization for what the judge wants to do"---at least if what they want to do would otherwise be morally repugnant. 

This is an emerging, and important area of research: how end users react to and use recommendations should be studied (and potentially modeled), and is central to the question of whether or not a model is effective and successful in a way far more meaningful than what can be captured by a metric \citep[as per the critique of][]{wagstaff2012}. 

\subsection{Alternatives: Modeling causality?}\label{sec:cause}
What is the alternative? I have frequently referred to a ``causal model'' as though this is possible, but the difficulty and uncertainty of this goal is perhaps the leading reason why we might want to find ways of using correlations alone. Of course, inferring causal relationships from correlations alone is unjustified \citep{freedman2005,freedman1997}; further tools are needed. One option is to return to statistical modeling, and specifically, the kind of causal inference done in econometrics. Insofar as this is possible, it addresses many problems in prediction-only approaches, but it has its own host of problems of statistical limitations and conceptual inconsistency, as summarized in a review by \citet{syll2018}. One problem is the ``fundamental problem of causal inference'' \citep{holland1986}, that we can never observe the same unit undergoing more than one treatment: so we need to make assumptions about how two units with different treatments are comparable (such as random assignment, or matching, or using structural relationships that can be interpreted as natural experiments such as with instrumental variables or regression discontinuity). This also means that cross-validation is a poor way of validating causal estimates \citep{wager2018}. 

\citet{gelman2009} notes that doing observational causal inference using natural experiments is limited to what natural experiments econometricians manage to find. And, even perfect application of techniques can break down in the presence of a latent confounder. This is the case in a fantastic example by \citet{arceneaux2010}, who show that the seemingly innocuous confounder of ``reachability by phone'' in a test of a get-out-the-vote campaign, is so powerful that it defeats perfect nonparametric matching on all of a rich set of observed covariates (including age, gender, party registration, and past voting, all which fail to control for reachability) in a sample size of over a million. 

The language of graphical models and do-calculus \citep{pearl2009} are a great advance in the ability to make precise causal statements, analyze relationships between such statements, and connect causality to probability distributions and statistical inference. But as a basis for \textit{estimating} causal graphical structure, \citet{humphreys1996} argue that the guarantees rely on strong, untestable, and probably false assumptions. More recently, \citet{freedman2004} remained skeptical about the assumptions needed to do causal learning, noting that substantial prior causal knowledge is needed for these models to work, but there are few causal pathways that can be excluded \textit{a priori}. 

And even as a representational tool, \citet{richardson2020} notes that causal graphs are capable only of representing short-term, immediate causal relationships; diffuse, long-term processes like structural racism \citep{richardson2016} don't fit easily into a directed acyclic graph. Similarly, as discussed above \citep{hu2019a,hu2020}, treating race or sex as variables in a directed acyclic graph forces them obey rules that directly contradict social science theory. If we rely on such representations, we risk constraining our thinking away from the solutions that are actually needed, just as \citet{barabas2018} talk about how prominent researchers in criminal justice risk assessment tools ``have explicitly discounted the relevance of sociological factors, such as ethnicity and socioeconomic status, because those factors are viewed as static and challenging to intervene upon.'' 

Where possible, experimental manipulations can introduce the full range of variability of inputs that allows observation of how they covary with outputs. But, even within the bounds of statistical theory, in order to interpret the coefficients of, say, a linear model as causal and generalizable, one would need \textit{random assignment} of treatments in a \textit{representative sample} of a target population \citep{aronow2016}\footnote{I thank \redact{Baobao Zhang} for this reference.} and an experimental setup which includes all \textit{key variables} (whether controlled or manipulated) without uncontrolled interference \citep{rosenbaum2007}. This may be demanding too much---for example, noncompliance \citepalias{student1931}, imperfect randomization, and participant crossover \citep{teele2014a} are persistent problems. On the latter point, omitting or controlling for some key variable or variables in an experiment could lead to a lack of ecological validity: that is, the causal processes that the experiment allows us to identify may not actually happen in the larger world. 

It is an open debate about how to balance these concerns, detailed in the collection edited by Dawn Langan \citet{teele2014}.\footnote{I thank \redact{Baobao Zhang} for this reference as well.} Randomized control trials are the ``gold standard'' for causal estimation, but without ecological validity and a representative sample (which is harder in random assignment), an estimate may mean relatively little. Another option is field experiments, which pursue greater ecological validity, but they sacrifice random assignment and the ability to control for all variables. 

Even with any type of experiment, there remain profound philosophical questions about whether causality, \citep[as defined in terms of manipulation, control, or change;][]{woodward2003} is knowable with any certainty: considering probabilistic notions of causality, is it sufficient that we observe that doing a certain manipulation produces a certain outcome in 80\% of cases to declare that we have found a causal relationship? What about 51\% of cases? And how can we ever truly rule out some hidden confounder as being the true cause? As in the thought experiment of Descartes' malicious demon, there could be a universal confounder that magically changes things to fool us into thinking there are causal relationships where there are none. We can also interpret this as what a ``black swan'' event does (treating it as something more specific than just an extremal value in a distribution): reveal some causal process that we did not know (or assumed away) beforehand, changing everything we thought we knew about a system. 

This also relates to the earlier critiques of quantitative versus qualitative analysis I gave; things that are difficult or impossible to quantify are just as important as things that can be quantified, creating blindspots in modeling, whether that modeling is causal or predictive. It also relates to the central tendency. Causal estimates being of average treatment effects again are of a central tendency; treatment effects may be very heterogenous within a population \citep{deaton2014}, and preferencing the average may disadvantage those who fall far from it. Estimating heterogenous treamtent effects partially ameliorates this and is superior insofar as it may be possible, but still is fundamentally only making finer and finer bins. Probabilistic models can say nothing once a bin has only one observation.

\section{Assessing model performance}\label{sec:CV}
Machine learning can claim to avoid all issues of interpretation, and of construct validity, internal validity, criterion-related validity \citep{devellis2017}, and even face validity, if it can demonstrate \textit{external validity}: that is, generalizability \citep[although the argument of][is precisely that these issues cannot be ignored for achieving certain gaols]{jacobs2019}. Within academic research, machine learning does so almost exclusively with the use of cross-validation: splitting data into a training set and a test set, and using performance on the test set as the basis for claims about performance on out-of-same data, i.e. about generalizability. Thus, the ability of cross-validation to establish generalizability is key for the overall validity of current machine learning research. 

\citet{rescher1998} writes about how each prediction necessarily entails a metaprediction---a prediction about how likely is that prediction to hold---and that systematically exploring this is a critical part of making predictions. Here, cross-validation gives ``predictions of the prediction error'' (or, estimates of the generalization error), and so is a metapredictive exercise, but we can also investigate one level higher: how likely is it for cross-validation to hold? 

There are a few distinct processes under which cross-validation can fail to hold. There are, of course, upstream problems. Bad labeling schemes, as discussed above, make everything following it meaningless. Filtering only to observations that have an observable signal \citep{tufekci2014} also makes models not generalizable, as in \citet{cohen2013}; in their case, the showed how going from standard filtering practices to a (more) representative sample led to classifier performance dropping from over 90\% to barely 65\%. If there is a distribution shift coming from the causal structure of the system that introduces never-before-seen covariance, as in Google Flu Trends, correlation-based modeling alone will fail. Not maintaining strict separation of the test set (e.g., extracting features before data splitting) can ``contaminate'' evaluations of performance \citep{domingos2012}. But beyond those problems, with cross-validation \textit{itself}, there are three major threats to validity:
\begin{enumerate}
    \item[\ref{sec:pub}] Publication bias.
    \item[\ref{sec:over}] Overfitting to the test set. 
    \item[\ref{sec:dep}] Dependencies among observations. 
\end{enumerate}

\subsection{Publication bias}\label{sec:pub}
\citet{gayoavello2012} notes that the ``predictions'' of machine learning research in social media are really ``claim[s] that a prediction could have been made, but the analysis is post hoc. And needless to say, negative results are rare.'' By this, he is referencing how negative results are not published, such that some positive findings may be the result of luck rather than something generalizable. That is: just because machine learning doesn't use $p$-values doesn't mean it escapes the problems of multiple comparisons: that, when making enough comparisons, by chance some uncorrelated variables will seem correlated. This also ties into the increasing attention to a ``reproducibility crisis'' in machine learning \citep{hutson2018}, just like in biomedical science or psychology, with publication bias being a concern alongside inaccessible data, undocumented tuning parameters and random seeds, and unusable code.

\subsection{Overftting to the test set}\label{sec:over}
Overfitting to the test set is a recognized problem \citep{rao2008,cawley2010,reunanen2003,dwork2015b}, but perhaps under-appreciated in larger discourse. \citet{dwork2015} write, ``in practice even data allocated for the sole purpose of testing is frequently reused... Such abuse of the holdout set is well known to result in significant overfitting on the holdout or cross-validation set.'' They give Kaggle competitions as an example of this. Kaggle is cleverly set up: competitions have a ``public'' leaderboard, where contestants can see their rankings through the duration of the competition, but also a ``private'' leaderboard, the results of which are not revealed until the end, and rankings on which are the basis for actually determining winners. The existence of this second, private, test set allows Kaggle to catch overfitting to the public test set. Kaggle forums contain discussions of people finding their ranking dropping enormously from the public to private leaderboard,\footnote{``What is up with the final leaderboard?'' \textit{Kaggle}, StumbleUpon Evergreen Classification Challenge. \url{https://www.kaggle.com/c/stumbleupon/discussion/6185}.} and Greg \citet{park2012} details how repeatedly submitting to rise on public leaderboards actually led to him unknowingly falling on the private leaderboard. 

In \textit{any} case where data is re-used for purposes other than testing, it can bias performance evaluation \citep{cawley2010}. This is a drawback to benchmark data sets; while it is incredibly important to have common data and measures of success to make sure that competing methods are compared on the same grounds, after a certain point any improvements to the state of the art may be overfitting \citep{gijsbers2019} \citep[another drawback is that a ``hyper-focus'' on benchmark sets comes at the expense of interpreting results in a domain context, and of making results that are actually applicable to a variety of uses;][]{wagstaff2012}. This is a place where significance testing may be good practice machine learning; McNemar's test \citep{mcnemar1947} is a version of the $\chi^2$ test for contingency tables applicable to paired comparisons, and therefore appropriate for the confusion matrices used to determine performance in machine learning \citep{dietterich1998}. This test is one way \citep{raschka2018} to test if an improvement in accuracy of one classifier versus another (or over a baseline of the majority class rate) is greater than might be expected by random chance, with other work offering appropriate tests for other metrics like precision \citep{gondara2016}. However, we should also make a correction for multiple comparisons; but without knowing how many attempts have been made across all researchers to improve on a certain baseline, we would not know what correction to make (or, equivalently, what would be a scientifically appropriate level at which to test). 

The third problem, dependencies, is my focus here, as it has received no systematic attention. First, I set up the context by introducing the idea within statistical theory of optimism, and then present an extension of the argument to connect to the issue of dependencies among observations.\footnote{Depending on the area, this problem is also known as non-iid data, autocorrelation (for certain types of dependencies), endogeneity, and `pseudoreplication.' I thank \redact{Jeremy Koster} for pointing me to the literature on pseudoreplication.} 

\subsection{Dependencies}\label{sec:dep}
In statistics, the problems of data that are not independent and identically distributed (iid) are well-known; dependencies shrink standard errors, and potentially cause bias in estimates. In causal inference and even experimental design, dependencies create interference \citep{rosenbaum2007} that can confound causal estimates. 

Machine learning, in not doing statistical inference, does not need to be concerned with estimating standard errors at all, and does not care about unbiased estimators; but it does not escape the consequences of the fundamental need of probability-based modeling to have multiple observations in order to accomplish anything. If machine learning fails to understand how dependencies change the effective sample size and affect modeling, it too suffers.  

Trivially, if the test set is an exact copy of the training set, obviously the performance on the test set is unreliable as a guide to out-of-sample performance. But duplicated observations are only the most simple case of dependent observations; partial dependencies will still affect training error. 

The main review of cross-validation by \citet{arlot2010} has only a cursory treatment of cross-validation with dependent data, and focuses on the one example of dependencies of time-series. Solutions that are appropriate mainly for that setting, such as ``modified CV'' \citep{chu1991}, similar to the $hv$-block cross validation of \citet{racine2000}, require that observations are uniquely ordered in sequence, that we have a way of characterizing distances between observations, and that there is a certain distance beyond which pairs are independent. While the general principle is good, the case of observations being ordered is fairly limited one. And even though probabilistic graphical models have a rich language for describing dependencies, it is not extended to metaprediction: the canonical textbook \citep{koller2009} does not connect these dependency structures to how to do data splitting for cross validation.

As a way of understanding this in more generality, I propose using the framework of \textit{optimism}, defined as the amount by which the training error---the empirical loss of the model under the same data used to fit the model---departs from the true, out-of-sample error \citep{efron2004}. 

There is a classic argument, for why the expected optimism will be strictly positive, making the training error less than the true out-of-sample error (and hence, `optimistic'). 

We imagine a data-generating process $Y \sim f$ with $\E_f(Y) = \mu$ and $\Cov_f(Y, Y) = \Var_f(Y) = \Sigma$. We use a draw from this process, $\y \in \R^n$, to form an estimator $\widehat{\bmu}_{\y}$ (perhaps using correlations between $\y$ and a set of features $\X$, not included here in the notation). For a set of unknown values $\y^*$, the predicted values would be $\widehat{\bmu}_{\y}(\y^*) = \widehat{\y}^*$. We can also apply this estimator back to make predictions of the same $\y$ that produced it, and we call these the \textit{fitted values} of $\y$,\footnote{These may also be called the `predicted values,' but I believe it makes sense to say `fitted values' if and only if talking about training/in-sample data.} $\widehat{\bmu}_{\y}(\y) := \widehat{\y}$. For theoretical analysis, we can analyze $\hat{\mu}_Y(Y) = \widehat{Y}$ as a random quantity, representing the \textit{procedure} of transforming a draw of $Y$ into an estimator \citep{tibshirani2019}. 

While the bias-variance tradeoff can for the most part be generalized \citep{friedman1997} to arbitrary loss functions \citep[although things can get strange for 0/1 loss;][]{domingos2000}, the bias-variance tradeoff is most easily illustrated with squared error loss. Squared error loss is empirically calculated by $\tfrac{1}{n}\|\y - \widehat{\y} \|^2_2$. We can see this as an estimator, $\widehat{\err}$, of the \textit{expected} in-sample error or \textit{expected} training error, to which we can apply the standard bias-variance argument (involving introducing a positive and negative version of certain terms in order to use probability identities), given here in matrix form:
\begin{align}
\err(\hat{\mu}) &= \tfrac{1}{n} \E_f \| Y - \widehat{Y}\|^2_2 \\
&= \tfrac{1}{n} \left[ \E_f \|Y\|^2_2 + \E_f \|\widehat{Y} \|^2_2 - 2\E_f(Y^T\widehat{Y}) \right] \\
&= \tfrac{1}{n} \left[ 
\E_f \|Y\|^2_2 
+
\E_f \|\widehat{Y} \|^2_2 
- 2\tr\E_f(Y\widehat{Y}^T) 
 \right] \nonumber \\
&\quad+ \tfrac{1}{n} \left[ 
- \mu^T \mu 
- \E_f(\widehat{Y})^T\E_f(\widehat{Y}) 
+ 2 \mu^T \E_f(\widehat{Y}) 
 \right] \nonumber \\
&\quad+ \tfrac{1}{n} \left[ 
\hphantom{-} \mu^T \mu 
+
\E_f(\widehat{Y})^T\E_f(\widehat{Y}) 
- 2 \tr \mu \E_f(\widehat{Y})^T 
 \right] \\
&= \tfrac{1}{n} \left[ 
\tr \Sigma 
+
\| \mu - \E(\widehat{Y})\|^2_2 
+
\tr \Var\nolimits_f(\widehat{Y}) 
-
2 \tr \Cov\nolimits_f(Y, \widehat{Y}) 
\right]\label{eq:cov1}
\end{align}
The error of an estimator $\widehat{\mu}$ in the \textit{test} set can be seen, again, as itself an estimator of the \textit{out-of-sample} error, also called the generalization error, the prediction error, or the test error. This quantity is defined as the error on a new draw from the data-generating process, $Y^*$, and is
\begin{align}
\Err(\hat{\mu}) &= \tfrac{1}{n} \E_f \| Y^* - \widehat{Y}\|^2_2 \\
&= \tfrac{1}{n} \left[ \tr \Sigma + \| \mu - \E(\widehat{Y})\|^2_2 + \tr \Var\nolimits_f(\widehat{Y}) - 2 \tr \Cov\nolimits_f(Y^*, \widehat{Y}) \right] \label{eq:cov2}
\end{align}
The only difference between the expected training error and the true out-of-sample error is in the respective covariance terms. $\widehat{Y}$ is dependent on $Y$ since it is formed from it (e.g., see eqn. \ref{eq:lm}), so the covariance term in eqn. (\ref{eq:cov1}) is nonzero. On the other hand, $\widehat{Y}$ is formed without any knowledge of $Y^*$, so they should be independent; and if they are, the covariance term in (\ref{eq:cov2}), $\tr \Cov\nolimits_f(Y^*, \widehat{Y})$ is zero. Then, the difference between the training and test error is called the \textit{optimism} \citep{efron2004}:
\begin{equation}
\mathrm{Opt}(\hat{\mu}) = \Err(\hat{\mu}) - \err(\hat{\mu}) = \tfrac{2}{n} \tr \Cov\nolimits_f(Y, \widehat{Y}).
\end{equation}
This is the expected amount by which the training error will under-estimate the true error.\footnote{While no general proof exists that the trace of this covariance is always positive (i.e., that optimism is always positive), \citet{zhang2004} shows that this only happens if either we use a nonconvex loss function (i.e., so the estimator is \textit{rewarded} the further it is from the training values), or if the estimator has \textit{negative} sensitivity to each data point (the more a given data point changes, the less the estimator changes; this is equivalent to the estimator having negative degrees of freedom). \citet{tibshirani2019} note that they cannot think of a way an estimator could have negative degrees of freedom, but they do not rule out the possibility such as in some pathological case.} For a linear smoother, e.g. $\bH \y$, this covariance term reduces to $\tfrac{2}{n}\tr \bH$; thus, an alternative to cross-validation is to estimate this covariance penalty directly \citep{efron2004}, through parametric or nonparametric means. 

Note that recently, \citet{rosset2019} have extended the analysis of optimism to the random-$X$ case, more appropriate to observational data, whereas previously a simplifying fixed-$X$ assumption has been made.\footnote{I thank an anonymous reviewer for pointing me to this work.} They show that randomness in $X$ adds an additional bias term and an additional variance term to the above derivations, although they make the iid assumption so it is possible that relaxing that assumption would also produce an additional covariance term. But it is unclear that, even if there is an additional covariance penalty, if this would suggest anything different for how to think about designing cross-validation procedures to guard against optimism in the test set. This is an opportunity for future investigation.

Dependencies between the training set and test set means that the covariance term in the optimism may not be zero, making the test \textit{set} error a downwardly biased (overly optimistic) estimator of the true out-of-sample error. In this sense, defining the ``test error'' as synonymous with the generalizability error may be unwise. 

I provide two examples of how this might happen. 

\subsection{AR(1) time series}

Consider a homoskedastic, mean-zero, stationary first-order autoregressive process, $Y_{t} = \phi Y_{t-1} + \varepsilon_{t}$, with $| \phi | < 1$ and $\bepsilon \sim \cN\left(\0, \sigma^2 \I\right)$. A 2-nearest neighbor regression estimator would, before data splitting, be $\hat{\mu}_Y(Y_t) = \widehat{Y_t} = \tfrac{1}{2}\left(Y_{t-1} + Y_{t+1} \right)$. If we randomly partitioned the data set into training and test, and the point $Y_t$ happened to fall into the test set whereas $Y_{t-1}$ and $Y_{t+1}$ were both in training, then the decomposition of the test error at point $t$ would include a term for the covariance between the point $Y_{t}$ and the point $\widehat{Y}_{t}$ (rather than between some independent copy $Y^*_{t}$ and $\widehat{Y}_{t}$). The autocovariance function of a mean-zero AR(1) process is $\gamma(h) = \E(Y_{t + h} Y_t) = (\sigma^2 \phi^{|h|})/(1 - \phi^2)$ \citep{shumway2011}, and the $t$-th diagonal element of the covariance matrix will be the average of the autocovariance between two sets of adjacent points:
\begin{align*}
    \Cov(Y_t, \widehat{Y}_t) &= \tfrac{1}{2} \E(Y_t(Y_{t-1} + Y_{t+1})) \\
    &= \tfrac{1}{2} \left[ \gamma(1) + \gamma(-1) \right] \\
    &= \tfrac{\sigma^2 \phi}{1 - \phi^2}.
\end{align*}
If $\phi$ is positive, then the test set error would be providing a downwardly biased, overly optimistic estimate of the true out-of-sample error. (Alternatively, if $\phi$ were negative, then the test set would provide an overly pessimistic, and upwardly biased estimate of the true out-of-sample error; but also, a 2-nearest-neighbor regression would not be a very good estimator.) A 4-nearest neighbor estimator would similarly give, for a held-out point $Y_t$ amidst four training points, the average of the covariances of two sets of adjacent points and two sets of points once removed, $\sigma^2 (\phi + \phi^2)/(1 - \phi^2)$, and so on.\footnote{Of course, what counts as nonzero optimism depends on what the ``generalized'' application is; if we had an imputation problem, random cross-validation would not be overly optimistic in its estimate of error. And for a one-ahead forecasting task, the whole point would be to know how well forecasts can do by leveraging any autocorrelation that is present, perhaps through a 2-nearest-left-neighbor estimator.} 

For forecasting, randomly partitioning a time series into training and test has, in terms of test set error serving as an assessment of real-world performance, been critiqued as ``time-traveling.''\footnote{This comes from the pseudonymous critique of the paper ``Twitter mood predicts the stock market'' and similar work in ``No limits to garbatrage,'' \textit{Buy the Hype} blog, August 29, 2013, \url{https://sellthenews.tumblr.com/post/59720892780/no-limits-to-garbatrage}.}

In a series of simulations, \citet{bergmeir2018} consider different cross-validation schemes on an autoregressive series. They consider $k$-fold cross-validation, ``non-dependent cross validation'' where partitions do not include adjacent time periods (resulting in many observations left out), leave-one-out cross validation, and temporal block cross-validation \citep[which they call ``out of sample,'' although as we saw for Google Flu Trends in][such blocks may still fail to correspond to out-of-sample performance for other reasons]{ginsberg2009}. For a linear fit to a low-order autoregression series, temporal block cross-validation gives a good estimate of error, whereas $k$-fold and LOO cross validation schemes give error rates half as high (non-dependent CV gives too-large estimates of error, probably because it has less data available for fitting). 

\subsection{Linear regression}\label{sec:lr}
In a ordinary least squares fit, the predicted values for an unknown $\y^*$ are made from the corresponding feature values $\X^*$ and weights estimated from known $\y$ and corresponding feature values $\X$. 
\begin{equation} \label{eq:lm}
\widehat{\bmu}_{\y; \X}(\y^*; \X^*)= \X^* (\X^T \X)^{-1} \X^T \y = \widehat{\y}^* 
\end{equation}
When we use $\X$ for $\X^*$, we can express the fitted values as $\widehat{\y} = \bH \y$ using the hat matrix $\bH = \X (\X^T \X)^{-1} \X^T$. 

Normally, we assume $Y \sim \cN (\X \bbeta, \sigma^2 \I)$. But now let us assume instead that all observations have a correlation $\rho$, and $Y \sim \cN (\X \bbeta, \bSigma)$, where $\bSigma_{ii} = \sigma^2$ and $\bSigma_{ij} = \rho \sigma^2$ for $i \ne j$, $0 < | \rho_{ij} | < 1$, and $\X \in \R^{n \times (p+1)}$ has a column of all ones. 

In order to both keep $Y$ random and to simply the algebra comparing across training and test sets, take the model 
\begin{equation} \label{eqn:N}
\begin{bmatrix}Y_1 \\ Y_2 \end{bmatrix} \sim \cN \left( \begin{bmatrix} \X \\ \X \end{bmatrix} \bbeta, \begin{bmatrix} \bSigma & \rho \sigma^2 \1 \1^T \\ \rho \sigma^2 \1 \1^T & \bSigma \end{bmatrix} \right).
\end{equation}
That is, let there be two exact copies of the feature levels $\X$, what \citet{rosset2019} call the ``fixed-X'' setting. Then we can assign observations $1,...,n$ to the training set and observations $n+1,...,2n$ to the test set, such that the $i$-th training observation and the $i$-th test observation will have the same mean, $\x_{i}^T\bbeta$. Given $Y$, the OLS fitted values for the training set are $\X (\X^T \X)^{-1} \X^T Y_1 = \bH Y_1$. Then, the optimism of the training set error is:
\begin{align}
\tfrac{2}{n} \tr \Cov\nolimits_f(Y_1, \widehat{Y}_1) = \tfrac{2}{n} \tr \Cov\nolimits_f(Y_1, \bH Y_1)  = \tfrac{2}{n} \tr \bH\Var\nolimits_f(Y_1) = \tfrac{2}{n} \tr \bH \bSigma 
\end{align}
But the test set error, too, is overly optimistic, with optimism:
\begin{align}
\tfrac{2}{n} \tr \Cov\nolimits_f(Y_2, \widehat{Y}_1) = \tfrac{2}{n} \tr \Cov\nolimits_f(Y_2, \bH Y_1) = \tfrac{2 \rho \sigma^2}{n} \tr \bH \1 \1^T = 2\rho \sigma^2,
\end{align}
with the last step coming from how the row sums of a hat matrix are all one,\footnote{In a with-intercept model, the first column of $\X$ is all ones, and $\bH$ is the orthogonal projection onto the column space of $\X$. Hence $\bH \1 = \1$} and the hat matrix times a matrix of all ones is a vector of its row sums repeated column-wise. 

Then, the other components of the expected out-of-sample error are the variance of the estimator, $\tfrac{1}{n} \tr \Var_f(\widehat{Y}) = \tfrac{1}{n} \tr \bH \bSigma \bH^T$, and the irreducible error, $\tfrac{1}{n}\tr \bSigma = \sigma^2$. For the test set error, the contribution of dependencies would be comparable in magnitude to the contribution of the irreducible error (e.g., for $\rho = .5$, the irreducible error would be seemingly canceled out, and as $\rho$ approaches 1, the irreducible error would seemingly be \textit{subtracted} from the test set error as opposed to added), making the test set error a biased estimator of the true out-of-sample error. Alternatively, if $\rho$ were negative (observations are anticorrelated), the test set error would be \textit{upwardly} biased, i.e. an overly \textit{pessimistic} estimator of the true out-of-sample error (the expected error on a new draw of data from the same distribution). But for a multivariate normal distribution of size $n$ with uniform diagonal elements and uniform off-diagonal elements, the correlation $\rho$ is constrained below by $\rho \geq -1/(n-1)$ to ensure the variance-covariance matrix only has positive eigenvalues and the distribution is well-defined. (Intuitively, as $n$ rises, it is less and less possible for all the dimensions of a multivariate normal to mutually have negative linear correlation.) So, negative correlation leading to possible pessimism is not a significant possibility, and does not balance concerns around possible optimism. 

\begin{figure}[!ht]
\centering
\includegraphics[scale=.75]{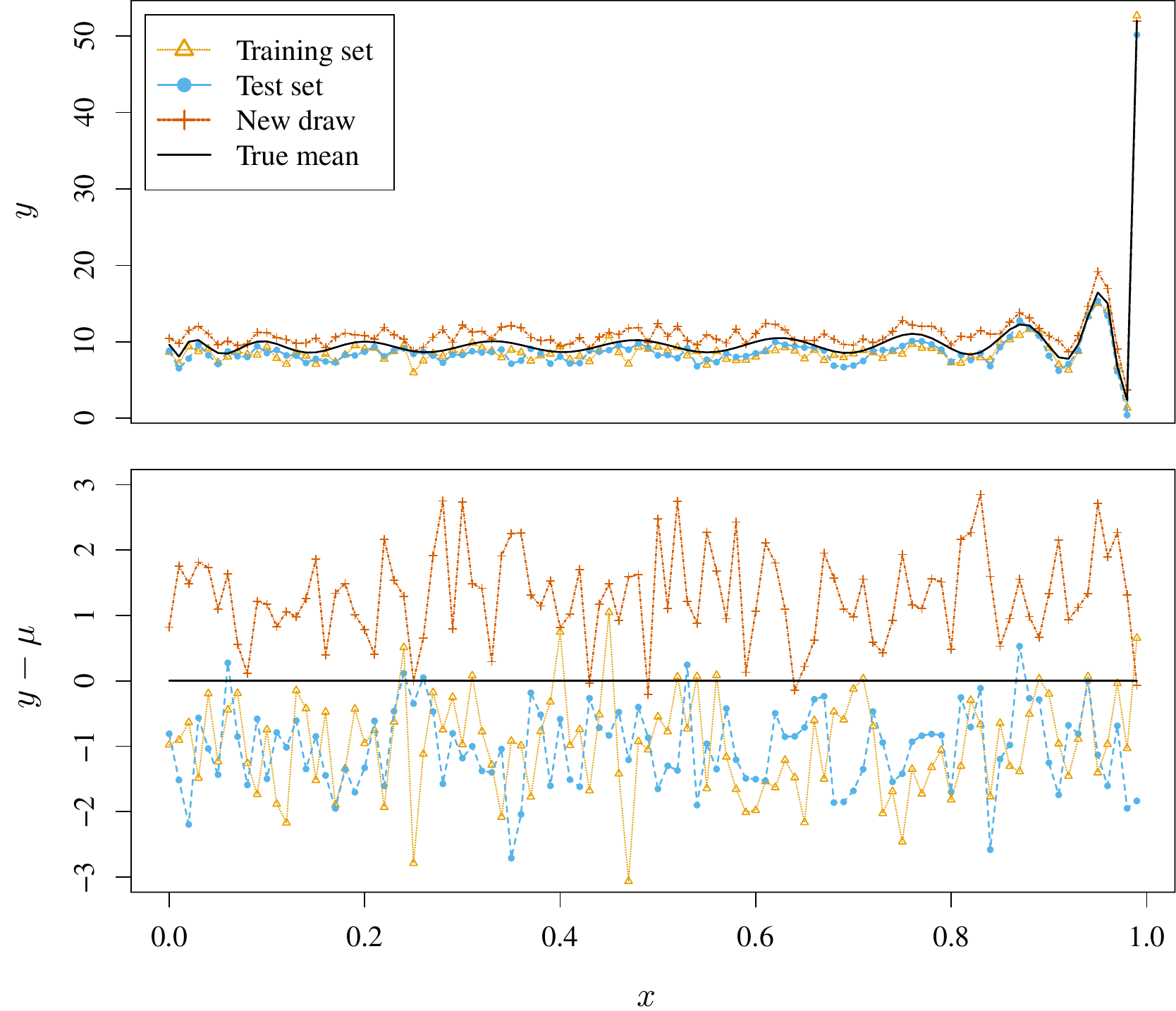}
\caption{Top: the curve generated from polynomial features, along with one draw from eqn. (\ref{eqn:N}) used as a training/test split, and an independent second draw. Bottom: for greater visibility, a plot of the difference between the draws and the true curve. Correlated observations can produce training and test sets close to one another but potentially far from an independent draw. Note that while the correlations between observations in one draw may bias a particular estimate, OLS is still an unbiased estimat\textit{or} because, across draws, the average bias is still zero.}\label{fig:data}
\end{figure}

Comparing the magnitude of optimism to the variance of the estimator is more difficult. Since $\bSigma$ is not diagonal, $\tr \bH \bSigma \bH^T = \tr \X (\X^T \X)^{-1} \X^T \bSigma \X (\X^T \X)^{-1} \X^T $ does not reduce to $\sigma^2 \tr (\X^T \X)^{-1}$ as usual, and it is harder to compare the magnitude of this term and determine the impact of the non-zero off-diagonals without having a specific form of $\X \in \R^{n \times (d+1)}$. But, as an example, if $\X$ is a matrix of uncorrelated features,\footnote{E.g., as achieved by orthogonal polynomials, where orthogonality for the $j$th and $k$th polynomials are defined as $\sum_{i=1}^n p_j(x_i) p_k(x_i) = 0$ and $\sum_{i=1}^n p_j^2(x_i) \neq 0$, and hence the correlation of polynomial feature $j$ and polynomial feature $k$ will be zero. The algorithm for computing such polynomials implemented in \textsf{R} is given in \citet{kennedy1980}.} then $\sigma^2 \tr (\X^T \X)^{-1} = \sigma^2 (d + \tfrac{1}{n})$, but $\tr \bH \bSigma \bH^T = \sigma^2 \left((1 - \rho) (d + 1) + \rho n \right)$, i.e. the variance of the estimator increases with the sample size proportionally to the amount of correlation between observations. This is reasonable, considering how dependencies might ``fool'' an estimator (consider the variability of a model that is fit to draws like those shown in fig. \ref{fig:data}). 

Important to note is that, in a multivariate normal response with nonzero positive correlation, \textit{there is no cross-validation scheme that would not lead to an under-estimate of the true error.} More generally, allowing arbitrary variance-covariance structure (i.e., $\bSigma_{ij} = \rho_{ij} \sigma_i \sigma_j$), and also covering the case of a no-intercept model, it would be difficult to say what would happen. The entries of the hat matrix are constrained, with positive diagonal entries generally larger in magnitude than off-diagonal entries \citep{mohammadi2016}, but there are negative off-diagonal entries that could interact with arbitrary correlations in arbitrary ways. Then, $\X$ could first be allowed to be random, but across a fixed set of values (like $\X$ being integer-valued within a given range) such that the levels of covariates may be matched up exactly between training and test, what \citep{tibshirani2019} call the ``same-X'' setting, or vary completely (like $\X$ being real-valued rather than integer-values), what they call the ``random-X'' setting where covariate levels do not match up. Both of these complicate analysis considerably. Working out the non-iid results in the same-X and random-X settings is a task for future analysis, and may contribute to understanding the amount by which we can decrease the bias in test error by designing cross-validation schemes that respect dependency structure. 

Simulating from the above model (eqn. \ref{eqn:N}), with $\X \in \R^{100 \times (20+1)}$ being the orthogonal polynomial features \citep{kennedy1980} from degree 0 to degree 20 on $n = 100$ points equally spaced in the interval $[0,1)$ with gaps of $0.01$, $\bbeta = \mathbf{10}$, $\rho = .5$, and $\sigma^2 = 1$, we can see the effect of correlated observations. Fig. (\ref{fig:data}) shows what it looks like to have correlated observations split into training and test sets be potentially far from an independent draw.

\begin{figure}[!ht]
\centering
\includegraphics[scale=.75]{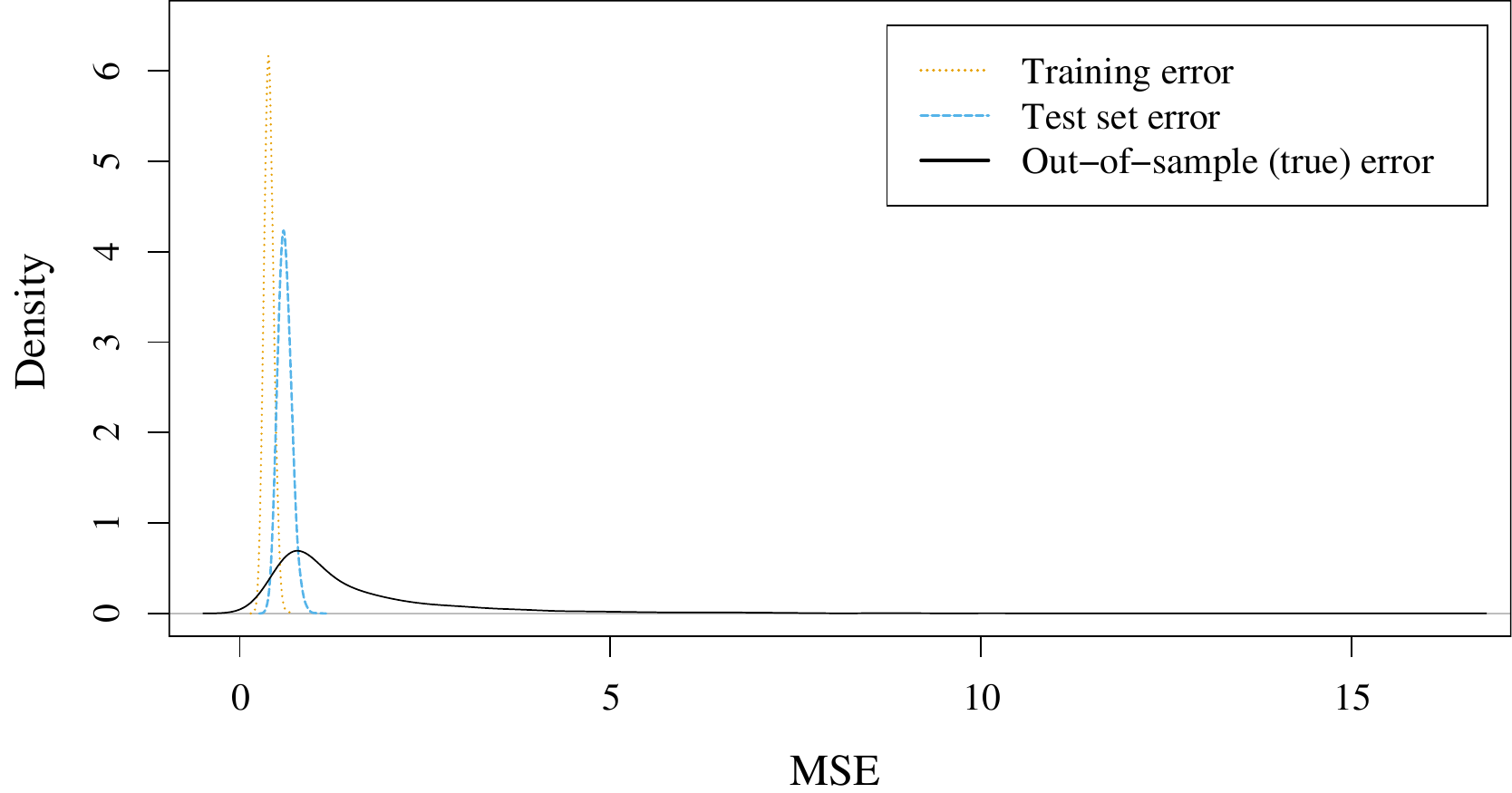}
\caption{Distribution of training error, test set error, and out-of-sample error over 10,000 replications (5,000 were not sufficient for the average out-of-sample to error converge to its expected value) for the model described in eqn. (\ref{eqn:N}).}\label{fig:mse}
\end{figure}

Fig. (\ref{fig:mse}) shows that this model indeed produces a true out-of-sample error higher in expectation than the average test set error. In particular, the out-of-sample error has only a slightly higher mode, but is right-skewed with large variance,\footnote{The mean squared error here is the OLS residual sum of squares.} leading to a substantially higher mean squared error than those of the training and test sets: $1.606$ versus, respectively, $0.396$ and $0.607$. Calculating from the above, we have:
\begin{itemize}
\item Irreducible error $\tr\bSigma/n = 1$;
\item Variance of the estimator $\tr \bH \bSigma \bH^T /n = 0.605$;
\begin{itemize}
\item Expected true out-of-sample error $1 + 0.605 = 1.605$;
\end{itemize}
\item Optimism in the training error $2 \tr \bH \bSigma/n = 1.21$;
\begin{itemize}
\item Expected training error $1.605 - 1.21 = 0.395$;
\end{itemize}
\item Optimism of test set $2\rho\sigma^2 = 1$; and
\begin{itemize}
\item Expected test set error $1.605 - 1 = 0.605$,
\end{itemize}
\end{itemize}
which agrees with the simulation to two significant figures. 

Raising $\rho$ towards $1$ makes the distribution of out-of-sample error flatten out even more and have an even longer tail; and, $\rho$ cannot be lowered more that slightly below zero (for $n = 100$, $\rho$ cannot be less than $-1/99$, around $0.01$) to keep $\bSigma$ positive definite, a point at which there is not much difference between the test set error and out-of-sample error. 

While it would not necessarily give insight into the relative contribution of dependencies to the MSE, it may be possible to derive the theoretical distributions that correspond to those shown in fig. (\ref{fig:mse}), based on results for quadratic forms \citep{mathai1992} for $\|Y_1 - \bH Y_1\|_2^2 = Y_1^T (\I - \bH) Y_1$, which follows a generalized non-central chi-square distribution \citep{mohsenipour2012}, and bilinear forms for $\|Y_2 - \bH Y_1\|_2^2$ and $\|Y^* - \bH Y_1\|_2^2$ \citep{mathai1992,mathai1995}. Unfortunately, there is no closed-form expression for the generalized non-central chi-square distribution, only an MGF; and similarly, for the bilinear forms, there are only expression for calculating cumulants. While these may give more detail about theoretical quantities relating these distributions, this is a fairly narrow example, so I avoid the exercise. 

\subsection{Consequences}
Based on what points are partitioned into a test set, from a known data-generating process we could compute the optimism of the test set error. Better yet, we could try to design cross-validation schema to minimize this optimism. 

Temporal autocorrelation, as in the first example above, is the simplest form of dependency.  To avoid ``time traveling,'' the best practice is to use temporal block cross-validation \citep{bergmeir2012}, where the test set is a block chronologically after the entire training set. More specifically, \citet{racine2000} recommends leaving out a block between the training and test sets in order to break the autocorrelation that would persists from having adjacent training and test blocks.

For other simple, regular forms of dependency, it is also possible to combat optimism in the test set error. \citet{hammerla2015} summarize how assessments of model performance in ubiquitous computing activity recognition has been led astray by dependencies that exist between multiple observations of the activity signatures of the same individual; in one case, test set error dropped from over 95\% to about 78.2\% when going from random data splitting to ``leave-one-subject-out'' validation. 

Network autocorrelation \citep{dow1982} gets more complex. If predictions are made about individuals who are connected in a network that causes network neighbors to be similar, having neighboring observations split across a training and test split will create optimism in the test set error. For example, label propagation for nodes on the boundary of any test/train split would have nonzero test set-estimator covariance structure very similar to that of nearest-neighbor regression and the AR(1) process described above. Even if an estimator did not use network features directly (or, indeed, if the network structure is unknown), the network structure may nonetheless make neighboring points more similar, making comparisons of neighbors across the train/test split unfairly positive, and we may not get a reliable picture of how well the estimator will generalize to new points outside of the observed network. 

If predictions are made about the network ties themselves, as in link prediction, there are nonetheless dependencies; network processes are dependencies \textit{between} ties, like reciprocity ($A_{ij}$ and $A_{ji}$), friend-of-a-friend connections ($A_{ij}$, $A_{ik}$, and $A_{jk}$), and cumulative advantage or preferential attachment ($A_{ij}$ and $A_{ik}$). For either node-level predictions or link prediction, if a network is partitioned into a training subgraph and a testing subgraph, the ties that are excluded are a means of ``sharing'' information between either the nodes or the ties of the training and rest sets. If there are no dependencies between ties other than, say, latent class membership of the nodes, then such a cross-validation scheme would be valid \citep{chen2018}; otherwise, the test set error undoubtedly suffers from optimism as well, perhaps unavoidably as in the non-independent multivariate normal case in the second example above. 

But in general, we would have to know or estimate the dependency structure to know the covariance between a test set and the estimator. And for a multivariate normal distribution with correlation as in the second example above, even if homoskedastic and with uniform correlations, there is no cross-validation scheme that would allow the test set error to be an unbiased estimate of the true out-of-sample error. If we knew we had such an intractable case, we would have to accept that there would be no avoiding some amount of optimism, or else we would need to collect additional data (rather than do data splitting). 

And as \citet{opsomer2001} point out, without assuming a parametric form for either the mean or covariance functions, models are not identifiable. So, \textit{no amount of additional data is ever enough}; without some independence assumptions about the structure of time, space, individuals, networks, or whatever else, additional observations cannot actually be used to estimate the covariance, because new observations themselves would have covariance that would need further observations to estimate \textit{ad infinitum}. 

And unlike wrong assumptions about, say, the model specification or functional form/function class (which machine learning can claim are irrelevant so long as performance is good), being wrong about the dependency structure can bias assessments of that performance itself. 

\subsection{Alternatives}
For making claims about machine learning systems, real-world testing is key. 

\subsubsection{Experimental testing}\label{sec:exp}
An excellent example of testing a machine learning system is given in \citet{cardoso2016}. This was based on an earlier paper, \citet{vantveer2002}, which trained a classifier (which, from the supplementary information, seems like a sort of custom-made decision tree, validated with leave-one-out CV) to find a set of gene expressions out of a set of 5,000 that correlated most highly with developing breast cancer. The 2016 paper took this system and subjected it to not only real out-of-sample testing, but did so in a randomized control trial to test the effect of relying on machine learning for decision-making. Specifically, patients were assessed for their ``clinical risk'' (traditionally made diagnosis) and their ``genomic risk'' (based on correlations found by the machine learning model). Patients who were high-risk on both criteria were always assigned chemotherapy, and patients who were low-risk on both criteria were never assigned chemotherapy. But those who were high-risk for one and low-risk for the other (the two tests disagreed) were randomly assigned to chemotherapy. The results: in cases where the ``clinical'' risk was low but the ``genomic'' risk was high, chemotherapy actually led to \textit{worse} outcomes! However, in cases where the ``clinical'' risk was high and the ``genomic'' risk was low, chemotherapy had no effect on outcomes, in which case avoiding a painful therapy is preferred. That is, \textit{relying only on the originally trained model alone for making chemotherapy decisions would have actually led to worse outcomes.} But, with the results of the experiment, as the authors point out \citep{cardoso2016}, we know how to use the model: as a second pass by which to catch false positives. Using it thusly, the study suggests, we will be able to avoid subjecting about 46\% of women with high clinical breast cancer risk to chemotherapy unlikely to be helpful. 

This study is not perfect: the original classifier was trained with the data of only 98 purposively sampled people, compared with the randomized control trial with 6,693 subjects (2,142 of which had discordant results to qualify for being randomly assigned). More data, representatively capturing as diverse a patient pool as possible, would expose the classifier to more variability over which to do better for non-majority cases. And, as large as was the sample for the experimental study---and done over 112 institutions in nine countries---all of those countries were European, potentially not sufficiently representing non-European populations. Testing should also explore how errors are distributed across subpopulations, as unequal distributions may advantage some groups less than others (or disadvantage some groups more); \citet{rodolfa2020} provide a good applied example of how to choose between and apply various proposed parity metrics. 

Then, rather than relying on a bespoke classifier written and trained in 2002, a better classifier might have been trained using tried-and-tested models in standard modern implementations, perhaps a random forest \citep{fernandez2014,caruana2008}. Even though requirements for and norms around medical testing are why there is even such a careful validation of the machine learning, the same regulations are likely part of what is behind this particular shortcoming of making continuous or even periodic updates very difficult. It took a decade for the original model to make its way through testing, and an updated model would need to be similarly validated before being approved for medical use. 

Ultimately, developing scientific understanding of which genes are causally linked to developing breast cancer would be a more solid basis on which to classify and then take action. But in terms of the use of machine learning itself, this is a good example of a meaningful, valid, and responsible approach, necessary (but not sufficient) for effective application. 

In terms of what else is necessary, beyond making sure that a given model to support treatment decisions is appropriate in technical terms, it should be seen, as phrased by \citet{sendak2020}, as ``a socio-technical system requiring integration into existing social and professional contexts.'' They theorize this as entailing the four key values and practices: ``rigorously define the problem in context, build relationships with stakeholders, respect professional discretion, and create ongoing feedback loops with stakeholders.'' Interpretability is not among these; and indeed, non-interpretable models whose reliability could be shown through rigorous testing would be perfectly consistent with other clinical practice, as ``the application of medical knowledge does not necessarily require the identification of causal relations. The human body is in many ways `a black box,' in which the causes and mechanisms of illnesses often elude explanation.'' They also note how prioritizing clinicians with ``substantial technical and quantitative expertise with which to engage with explainable machine learning'' will come at the cost of the emotional intelligence that is far more important for actually achieving effective care.

\subsubsection{Careful cross-validation schema}
For cross-validation, machine learning should begin to think systematically about how dependencies in data might be affecting the reliability of cross-validation, and develop and adopt cross-validation schemes to address this. Such efforts could draw on existing work on different forms of dependencies \citep{hammerla2015,bergmeir2012,chen2018}, mixing them and adding in new forms of dependencies. 

\subsubsection{Rhetorical change}
But true out-of-sample testing, and experimentally testing the result of \textit{intervening} on the basis of a result for machine learning, may not be possible, or preliminary results might be needed in order to determine whether such testing is worthwhile. In such a case, there are effective alternatives in terms of communicating the results of machine learning models. There are some basic issues of communications, such as in explaining the ``accuracy paradox.''\footnote{If the base rate is 1\%, a classifier that always guesses the majority class would be useless, but would achieve 99\% accuracy.} The technical meanings of precision and recall are not likely to be colloquially known, and conversely, the technical meaning of prediction in terms of correlations do not correspond to lay understandings of ``prediction,'' suggesting a strategy of giving an explanation of the meanings of these terms along with public presentation of results. 

The problem of over-optimism in assessments of model generalizability on the basis of cross-validation can similarly be addressed somewhat just by providing caveats about results being preliminary until experimental testing is done. Indeed, over-optimism is misleading not only to the public, but potentially to researchers as well, who may take their own performance claims as ``an assertion about the performance of the system under general conditions'' \citep{cohen2013} rather than as preliminary claims. Without understanding how out-of-sample testing is the ultimate arbiter, claims may enable the adopting of machine learning systems that turn out to not perform well. Worse, if there is no feedback mechanism for checking model performance in the wild (e.g., if human annotation was necessary for model building, and no further annotation is done once the system is deployed), we may never know that a model is failing to perform well. 

What are alternative to current practices? For publication bias, it might be pre-registration as has been adopted in other areas. For overfitting to test data, there are promising technical proposals \citep{dwork2015b}, but like any procedure there will be trade-offs we may not want to make and assumptions that might fail. A purists' solution would be to set aside data and use it for testing only \textit{once}. That one test is what must be reported; if the test performance turns out to be too low to be publishable (outside of a pre-registration database) or applied, for example, then the whole project must be abandoned (or else a new test set must be gathered). But this is likely neither feasible nor desirable. Held-out data is already wasteful in terms of getting better estimates; the goal of getting a good estimate of the out-of-sample error is important enough to make this sacrifice, but if held-out data was only usable once, it would be too high a price. Here I agree with \citet{irpan2018}: certain costs of reproducibility may be higher than its benefits. A better solution, then, is simply to be more honest about the likely extent of overfitting even with data splitting and cross-validation. 

Overall, research can acknowledge that statements about ``predictions'' from cross-validation are post-hoc statements about the strength of correlations \citep{gayoavello2012} that are only partially reliable guides to generalizability. \citet{rescher1998} writes, ``\textit{of course} past performance is a predictive indicator. (What could possibly serve better?) What past performance does \textit{not} enable one to do is predict with failproof accuracy.'' To emphasize this point and avoid colloquial confusion (and emphasize that machine learning predictions are based on past performance, and not on causal understandings), predictions are probably better called ``backtesting'' or ``retrodiction'' \citep[although the use of these terms alone is not sufficient to prevent modelers from overfitting, as][argue happens in financial modeling]{bailey2014}. The use of `prediction' to refer to fitted values is quite old and inherited from statistics, and so this is not an example of machine learning engaging in ``wishful mnemonics'' \citep{mcdermott1976} or misusing language via ``choosing terms of art with colloquial connotations'' \citep{lipton2018b}, but that does not mean that machine learning practitioners and researchers are not capable of changing this usage. 

\vspace{10pt}

In all cases, cross-validation evaluations of model performance should be taken not as a guarantee, but as preliminary. Only true out-of-sample testing, in a setting mimicking what an actual application would be (including, where appropriate or possible, experimental testing), is what can truly determine a machine learning model's success.

\section{Summary and conclusion}
In this paper, I have attempted to systematically review grounds for the ways in which machine learning models are limited and can fail as applied to social systems. The described hierarchy can be used both for challenging models, for example when they lead to unjust ends, and for modelers to understand the limitations of machine learning, in order to try and mitigate potential failures and to better communicate what modeling actually demonstrates and achieves. 

In summarizing the commitments of machine learning, I draw on work from Science and Technology Studies, theories of measurement, and critiques of statistics still applicable to the way machine learning sees the world. Summarizing work on the difference between prediction and explanation \citep{shmueli2010,breiman2001}, I make connections both to philosophy \citep{rescher1998} and to current debates about explainability and interpretability \citep{doshivelez2017,caruana2015,lipton2015}. Lastly, I bring together materials calling into question the reliability of cross-validation estimates of generalizability error, including unifying discussions of how to properly do cross-validation in a framework of non-iid data, demonstrating the effect on test set optimism and showing how issues identified by statisticians in the past can arise in new forms for machine learning. 

As mentioned in the introduction, systematically developing ways of integrating other methods into machine learning project pipelines, or even making true mixed-methods pipelines, represents the most promising way forward of overcoming limitations and failure points of machine learning-only research. At every point where machine learning can fail, there are alternatives that will not fail; conversely, places where other methods fail may be places where machine learning can provide a way forward. 

In terms of alternatives, my work suggests the following tasks:
\begin{itemize}
    \item Systematically incorporate qualitative methods;
    \item Seek ways of including variability, or higher moments, as one way of mitigating a reliance only on the central tendency corresponding to the first moment;
    \item Split data for cross-validation in ways that take dependencies into account;
    \item Do out-of-sample, real-world testing before making strong claims about model performance (and, conversely, demand such testing before accepting strong claims);
    \item Be aware, open, and humble about the limitations of machine learning modeling. 
\end{itemize}
That is, machine learning modelers should also consider how some tasks might be better addressed with statistics, other forms of mathematical modeling, or even qualitative research. 

But returning to the cautionary point in the introduction, mixed methods are not easy to do. A researcher trained in machine learning should not think there is some one book they can read and then be qualified to undertake an ethnography, or even to design survey questions, any more than the converse. Jason \citet{seawright2016}\footnote{I thank \redact{Baobao Zhang} for this reference.} points to another subtlety of mixed methods: specifically, he argues \textit{against} a parallel or linear ``triangulation'' approach to mixed methods, which tries to answer the same question with different methods in order to try and get convergence, noting that since ``qualitative and statistical approaches produce results that are different in kind, it is only possible to assess such convergence very abstractly.'' He instead advocates for an integrative approach of putting different methods in sequence (potentially iteratively), such as statistically estimating coefficients ``from a model designed, refined, and tested in light of serious qualitative analysis.'' Similarly, machine learning could try to train predictive models designed, refined, and tested in light of serious qualitative inquiry. Or, perhaps taking into account the history of ethical failing in qualitative research that was not based on reciprocal exchange \citep{smith2012}, models could be designed, refined, and tested in light of meaningful participation of human subjects (meaningful in the sense of having veto power, and not just serving perfunctorily as non-binding consultation) over what labels are used, for building which models, to accomplish what ends. \citet{mcquillan2018} proposes popular assemblages, in the model of patients' councils, as an example of the institutional form such input might take for machine learning. \citet{dunning2008}\footnote{I thank \redact{Baobao Zhang} for this reference as well.} also talks about the role of qualitative methods in the design of experiments, which we could also apply to experimental testing of machine learning like that of \citet{cardoso2016}. 

In the short term, attempts at machine learning that engage in good faith with the limitations of modeling are better positioned to identify and avoid pitfalls of unjust development and application. They will also avoid contributing to unrealistic public expectations that could harm the vitality of the field. A healthy skepticism in approaching modeling can therefore be a central part of developing a community of responsible practice. In the long term, developing frameworks, pipelines, best practices, and experience in actually incorporating alternative methods into uses of machine learning holds promise for reliably interacting with the world to improve the human condition, free of unintentional or unanticipated harms. 

\section*{Conflict of Interest Statement}
The author declares that the research was conducted in the absence of any commercial or financial relationships that could be construed as a potential conflict of interest.

\section*{Author Contributions}
Momin M. Malik did all work enclosed, with inputs from other individuals given in footnotes. 

\section*{Funding}
Funding was provided by the Ethics and Governance of AI Fund at the Berkman Klein Center for Internet \& Society at Harvard University. The views and conclusions contained in this document are those of the author and should not be interpreted as representing the official policies, either expressed or implied, of the Berkman Klein Center, the Ethics and Governance of AI Fund, or any other entity.

\section*{Acknowledgments}
Portions of this paper were presented at the 2019 Information Ethics Roundtable, and other parts are drawn from a 2019 blogpost. In addition to the in-line acknowledgements, I thank my IER respondant, Joshua Simons, for invaluable feedback, and Karthik Dinkar for the idea to do this piece in its specific form as well as for feedback. Thanks to Hemank Lamba, joint work with whom the review of the properties of the lasso were drawn. Thanks to Beau Sievers for fantastic feedback on an intermediate version the draft. 

Any errors, unfair generalizations and characterizations, or flawed reasonings are my own. 

\bibliography{ml_hierarchy_plain}

\end{document}